\documentclass[lettersize,journal]{IEEEtran}
\usepackage{amsmath,amsfonts}
\usepackage{algorithmic}
\usepackage{algorithm}
\usepackage{graphicx}

\usepackage{array}
\usepackage[caption=false,font=normalsize,labelfont=sf,textfont=sf]{subfig}
\usepackage{textcomp}
\usepackage{stfloats}
\usepackage{url}
\usepackage{verbatim}
\usepackage{graphicx,color,xcolor}
\usepackage{multirow}
\usepackage[colorinlistoftodos]{todonotes}
\usepackage{booktabs}
\usepackage{fdsymbol}

\definecolor{colortDCF}{RGB}{0, 103, 181}
\definecolor{colorEER}{RGB}{210, 72, 24}

\usepackage[
backend=biber,
style=ieee,
maxbibnames=3,
maxcitenames=3,
doi=false,isbn=false,url=false,eprint=false
]{biblatex} 

\defbibheading{bibliography}[\refname]{}

\begin{document}

\title{ASVspoof 2021: Towards Spoofed and Deepfake Speech Detection in the Wild}

\author{Xuechen Liu, Xin Wang, \emph{Member, IEEE}, Md Sahidullah, \emph{Member, IEEE}, Jose Patino, Héctor Delgado, Tomi Kinnunen, \emph{Member, IEEE}, Massimiliano Todisco, \emph{Member, IEEE}, Junichi Yamagishi, \emph{Senior Member, IEEE}, Nicholas Evans, \emph{Member, IEEE}, Andreas Nautsch, Kong Aik Lee, \emph{Senior Member, IEEE}
\thanks{Xuechen Liu and Md Sahidullah were with Universit\'{e} de Lorraine, CNRS, Inria, LORIA, F-54000, Nancy, France (e-mail: xuechen.liu@inria.fr, md.sahidullah@tcgcrest.org). Xuechen Liu is now with National Institute of Informatics, Tokyo 101-8430, Japan. Md Sahidullah is now with Institute for Advancing Intelligence, TCG CREST, India.}
\thanks{Xin Wang and Junichi Yamagishi are with the Digital Content and Media Sciences Research Division, National Institute of Informatics, Tokyo 101-8430, Japan (e-mail: wangxin@nii.ac.jp, jyamagis@nii.ac.jp).}
\thanks{Jose Patino, Massimiliano Todisco, Andreas Nautsch, and Nicholas Evans are with Digital Security Department, EURECOM (Campus SophiaTech), 06410 Biot, France (e-mail: jose.patino@cerence.com, todisco@eurecom.fr, evans@eurecom.fr). Jose Patino is now with Cerence Inc. Andreas Nautsch is now with Avignon Universit{\'e}.}
\thanks{Héctor Delgado is with Nuance Communications, C/ Gran Vía 39, 28013 Madrid,
Spain (e-mail: hector.delgado@nuance.com).}
\thanks{Tomi Kinnunen and Xuechen Liu are with School of Computing, University of Eastern Finland, FI-80101, Joensuu, Finland (e-mail: tomi.kinnunen@uef.fi).}
\thanks{Kong Aik Lee is with the Institute for Infocomm Research, A$^\star$STAR, 138632 Singapore (e-mail: lee\_kong\_aik@i2r.a-star.edu.sg).}

\thanks{This paper has supplementary material provided by the author. The material includes supplementary\_material.zip. The materials will be uploaded to \texttt{https://www.asvspoof.org/index2021.html}. Contact xuechen.liu@inria.fr for further questions about this work.}
\thanks{Published in IEEE/ACM-TASLP (doi: 10.1109/TASLP.2023.3285283) \textcopyright IEEE. Personal use of this material is permitted. Permission from IEEE must be obtained for all other uses, in any current or future media, including reprinting/republishing this material for advertising or promotional purposes, creating new collective works, for resale or redistribution to servers or lists, or reuse of any copyrighted component of this work in other works.}
}

\markboth{PUBLISHED IN IEEE/ACM TRANSACTIONS ON AUDIO, SPEECH, AND LANGUAGE PROCESSING (10.1109/TASLP.2023.3285283)}%
{Shell \MakeLowercase{\textit{et al.}}: A Sample Article Using IEEEtran.cls for IEEE Journals}

\maketitle

\begin{abstract}
Benchmarking initiatives support the meaningful comparison of competing solutions to prominent problems in speech and language processing. Successive benchmarking evaluations typically reflect a progressive evolution from ideal lab conditions towards to those encountered in the wild. ASVspoof, the spoofing and deepfake detection initiative and challenge series, has followed the same trend. This article provides a summary of the ASVspoof 2021 challenge and the results of 54 participating teams that submitted to the evaluation phase. For the logical access (LA) task, results indicate that countermeasures are robust to newly introduced encoding and transmission effects. Results for the physical access (PA) task indicate the potential to detect replay attacks in real, as opposed to simulated physical spaces, but a lack of robustness to variations between simulated and real acoustic environments. The Deepfake (DF) task, new to the 2021 edition, targets solutions to the detection of manipulated, compressed speech data posted online. While detection solutions offer some resilience to compression effects, they lack generalization across different source datasets. In addition to a summary of the top-performing systems for each task, new analyses of influential data factors and results for hidden data subsets, the article includes a review of post-challenge results, an outline of the principal challenge limitations and a road-map for the future of ASVspoof.
\end{abstract}


\section{Introduction}
\IEEEPARstart{B}{iometric} systems implemented with automatic speaker verification (ASV) technology \cite{Bai2021-speaker-rec-deep-learning-overview} are vulnerable to \emph{spoofing attacks}~\cite{ISOpresentationAtack} whereby an adversary attempts to masquerade as another individual through the presentation of manipulated speech data~\cite{wu2015spoofing}. 
There are four principal means to generate attacks: \emph{impersonation}, \emph{voice conversion} (VC), \emph{text-to-speech} (TTS) synthesis, and \emph{replay}. 
Impersonation attacks have received the least attention; successful attacks typically require special expertise, e.g.~of professional impersonators. In contrast, VC, TTS and replay spoofing attacks can all be mounted using readily available software toolkits and consumer devices and have hence received greater attention~\cite{wu2015spoofing}. The threat posed by such techniques is now well recognised, particularly in academia and increasingly in industry~\cite{VoiceID,Binsider,Pindrop,FacebookMicrosoft}.

The ASVspoof initiative and challenge series~\cite{Wu-ASVspoof2015} was conceived to foster the development of countermeasures (CMs) to protect against the manipulation of ASV systems from spoofing attacks.
ASVspoof has designed, collected and distributed substantial databases of both bona fide and spoofed data, with the latter being generated with a broad range of state-of-the-art VC and TTS technologies, recording and replay devices.
They have been used in the series of four biennial benchmarking challenges, the results of which show tremendous progress in spoofing detection reliability.

The first edition of ASVspoof in 2015~\cite{Wu-ASVspoof2015} focused on the development of CMs for the detection of VC and TTS attacks. For the subsequent edition in 2017, the focus switched to replay attacks. The third edition, ASVspoof 2019, was the first to address all three attack types through two separate tasks involving: (i)~a logical access (LA) scenario involving VC and TTS attacks generated using a set of different statistical and neural methods; (ii)~a physical access (PA) scenario involving replay attacks implemented in a large number of simulated acoustic environments. 

While the overall scope of ASVspoof challenges has been steadily expanding with the introduction of new attack algorithms, evaluation scenarios, and performance metrics~\cite{Kinnunen2020-tDCF-fundamentals}, the technical quality of the audio data to date has remained high; the data is relatively free of additive noise and encoding, compression and transmission artifacts. 
While experiments with clean data detached from the complexities of the real world have a legitimate role in basic research, the prolonged use of such data may promote the development of CM solutions which fail to generalize to application scenarios in the wild where external distortions are commonplace. Furthermore, the past ASVspoof training, development and evaluation sets all originate from the same source corpus: either VCTK~\cite{vctk} or RedDots~\cite{lee15_interspeech}.
This may hinder the design of CMs which generalize to other domains with different speech characteristics.

\begin{figure*}[th!]
     \centering
     \includegraphics[width=0.99\textwidth]{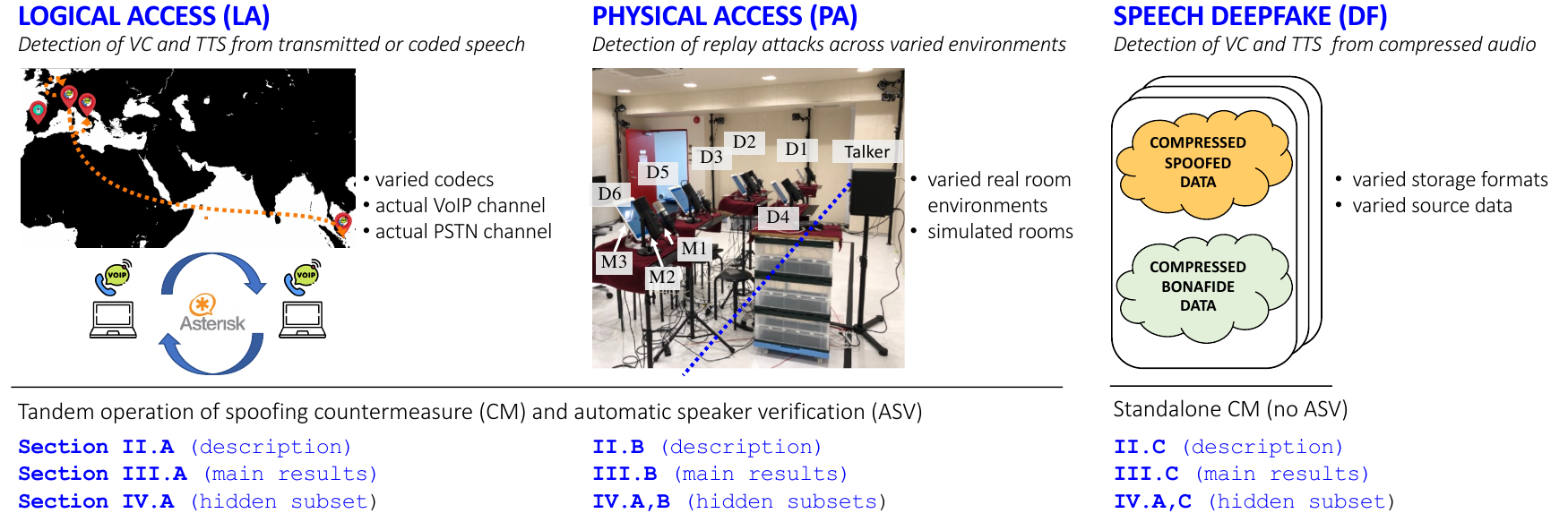}
     \vspace{-2mm}
     \caption{The ASVspoof 2021 challenge consists of three different tasks: logical access (LA); physical access (PA); deepfake (DF). 
     While the general theme of exposing spoofing countermeasures (CMs) to more realistic conditions is common to all three tasks, each was effectively run as a separate sub-challenge. The database and results for each task are described in the sections indicated.
     }
     \label{fig:ASVspoof21_overview}
\end{figure*}

As illustrated in Fig.~\ref{fig:ASVspoof21_overview}, the move towards an evaluation using speech data which is more representative of practical applications was the essence of the ASVspoof 2021 challenge.  
It sought to benchmark the latest CM solutions in more realistic conditions in which speech data undergoes coding, compression and transmission across telephony channels (LA task, left in Fig.~\ref{fig:ASVspoof21_overview}), or undergoes acoustic propagation in a real physical space (PA task, middle in Fig.~\ref{fig:ASVspoof21_overview}).
ASVspoof 2021 also introduced the new deepfake (DF) task (right in Fig.~\ref{fig:ASVspoof21_overview}) in which adversaries aim not to deceive an ASV system, but merely fabricate an utterance in the voice of a target speaker.  
The goal may be to harm the reputation of a well-known personality or to manipulate society by disseminating disinformation or fake news~\cite{DeepfakesNews1,DeepfakesNews2}.
The aim behind the new DF task is to assess the robustness of spoofing detection solutions when used to detect compressed manipulated speech data of varying characteristics posted online.
The scenario is simulated by processing audio files from different sources with various codecs used in social media. 

Described in this paper are the key advances made since previous editions of ASVspoof, together with details of the dataset design, collection and protocol policies.  
Also described are the challenge results and principal techniques that are common to the top-performing systems for all three sub-tasks. 
The study substantially extends the preliminary workshop version~\cite{yamagishi21_asvspoof} with an in-depth analysis of influential data factors that impact upon CM performance. 
Additionally, we present a summary of system descriptions submitted by participating teams and identify the most promising detection algorithms and techniques.
The article also contains a new, detailed analysis of results for specially-designed \emph{hidden subsets} and includes a survey of \emph{post-challenge} studies and related work outside of the scope of ASVspoof. None of these were presented in~\cite{yamagishi21_asvspoof}.  

The new insights presented in this paper, together with newly released meta-data for the three ASVspoof 2021 task databases\footnote{The meta-data was used in the post-evaluation analysis presented herein and is publicly available via the challenge website: \url{https://www.asvspoof.org/index2021.html}.} 
should be helpful to those looking to participate in future ASVspoof challenges as well as all those working in the field. 
Toward the end of the paper we reflect upon the limitations and key lessons learned from ASVspoof 2021, with a discussion of current ideas and directions for future research.

\section{ASVspoof 2021 Challenge Outline}
In this section, we outline the three challenge tasks, each an independent sub-challenge with their own corresponding evaluation database. All three are partitioned into progress and evaluation subsets in addition to a number of hidden subsets. Progress subsets of modest size were used for intermediate assessment prior to the evaluation submission deadline for which participants submitted scores for the full evaluation database comprising the three subsets. Only progress and evaluation subsets are discussed in this section, with corresponding results being presented in Section~\ref{sec:results}. Hidden subsets and results are presented in Section~\ref{sec:Hidden-Tracks}.
It is worth noting the difference between the progress set and development set: In the former, participants do not have access to the ground-truth speaker labels, while for the latter the speaker labels are available for tuning the training progress. Moreover, during the challenge, participants could make only a limited number of score submissions during the progress phase.

For all three tasks, all data was distributed in free lossless audio codec (FLAC) format with a sampling rate of 16~kHz and with 16-bit quantisation. There was no new training data for ASVspoof 2021, though protocols for training and development using data released through previous ASVspoof editions were made available. Participants were required to use LA training partition from ASVspoof 2019 for training CM systems for LA and DF tasks, and PA training partition from the same dataset for PA task. Challenge rules and participant guidelines are available in the challenge evaluation plan~\cite{delgado2021asvspoof}. 

\begin{table}[!t]
    \centering
    \caption{Number of trials and speakers in each task.}
      \begin{tabular}{|r|r|r|r|r|r|}
        \hline
              &       & \multicolumn{2}{c|}{ \#. trials} & \multicolumn{2}{c|}{ \#. speakers} \\
        \cline{3-6}
        Task & Subset & Bona fide & Spoofed & Female & Male \\
        \hline
        \hline
        \multirow{2}{*}{LA} & Progress & 1,676 & 14,788 & 37 & 30  \\ 
        &  Evaluation & 14,816 & 133,360  & 37 & 30 \\ 
        \hline
        \multirow{2}{*}{PA}  & Progress & 14,472 & 72,576  & 37 & 30\\ 
        & Evaluation & 94,068 & 627,264  & 37 & 30 \\ 
        \hline
        \multirow{2}{*}{DF} &        Progress & 5,768 & 53,557  & 37 & 30 \\ 
        & Evaluation & 14,869 & 519,059  & 50 & 43 \\ 
        \hline
    \end{tabular}
    \label{tab:All_trial_num}
\end{table}

\subsection{Logical Access (LA)}
\label{sec:LAdb}

The ASVspoof 2015 and 2019 editions both incorporated LA tasks in which all speech signals were clean, i.e.\ without either additive noise, reverberation or other sources of channel variation.
As a consequence, the research effort focused upon the identification of key techniques for the capturing and classification of processing artifacts which result from the generation of spoofed speech signals using either TTS or VC algorithms.
Such ideal conditions are unrealistic and other work has shown the likely degradation in CM performance when they are deployed in practical scenarios such as person authentication over telephony channels~\cite{Delgado_channel_variation2017,Lavrentyeva_phonespoof_2019}.

The ASVspoof 2021 LA task was hence designed to reduce the gap between ideal laboratory conditions and those to be expected in the wild. 
This was achieved with the transmission of both bona fide and spoofed speech across real telephony systems, including a voice-over-internet-protocol (VoIP) system and a public switched telephone network (PSTN). The 2021 LA challenge hence focuses on the study of robustness to nuisance variation from compression, packet loss and other artifacts stemming from different bandwidths, transmission infrastructures and bitrates, etc.; the consideration of additive noise remains for a future challenge edition. 

\begin{table}[t]
    \centering
    \caption{Summary of LA data conditions. \underline{Underlined factors} appear in evaluation subset but not in progress subset}
    \begin{tabular}{|c|c|c|c|c|}
        \hline
        Cond. & Codec	& \begin{tabular}{@{}c@{}} Sampling \\ rate \\ \end{tabular}   & Transmission & Bitrate\\ 
        \hline\hline
        LA-C1 & - & 16 kHz & - & 250 kbps\\ 
        LA-C2 & a-law & 8 kHz & VoIP & 64 kbps\\ 
        LA-C3 & unk. + $\mu$-law & 8 kHz & PSTN + VoIP & - / 64 kbps\\ 
        LA-C4 & G.722 & 16 kHz & VoIP & 64 kbps\\ 
         \underline{LA-C5} & $\mu$-law & 8 kHz & VoIP & 64 kbps\\ 
         \underline{LA-C6} & GSM & 8 kHz & VoIP & 13 kbps\\ 
         \underline{LA-C7} & OPUS & 16 kHz & VoIP & VBR ~16 kbps\\ 
        \hline
    \end{tabular}
    \label{tab:LA_conditions}
\end{table}

Speech data are sourced from the ASVspoof 2019 LA evaluation database, itself derived from the VCTK database~\cite{vctk}. 
Spoofed trials are generated using one of 13 different VC, TTS or hybrid spoofing attack algorithms (A07-A19 in~\cite{asvspoof2019database}).
For all but one condition, \emph{both} bona fide \emph{and} spoofed data undergo transmission across either a VoIP or PSTN+VoIP network using one of six  codecs giving the seven evaluation conditions listed in Table~\ref{tab:LA_conditions}.
The choice of codecs is broadly representative of traditional or legacy codecs (a-law, G.722, etc.) and modern IP streaming codecs (OPUS) in use today.

The reference condition C1 is identical to the ASVspoof 2019 LA scenario, i.e.\ with neither encoding nor transmission.
Conditions C2 and C4-C7 correspond to transmission across an Asterisk\footnote{\url{https://www.asterisk.org}} private branch exchange (PBX) system using codecs operating at sampling rates of either 8~kHz (C2: a-law, C5: $\mu$-law, C6: GSM Full Rate 6.10) or 16~kHz (C4: G.722, C7: OPUS).
The bitrate of each data condition (last column of Table~\ref{tab:LA_conditions}) was set commensurate with the codec, with the lowest being 13 kbps for condition C6.
Each utterance is transmitted in its own individual VoIP channel, by automatically generated Asterisk call files and between two session initiation protocol (SIP) endpoints. The transmission is done either within a local area network connection, or from an endpoint hosted in France to endpoints in Italy or Singapore.
Condition C3 involves transmission in Spain over a PSTN system for which codec conditions are uncontrollable and unknown.
Calls are initiated from a mobile smartphone whereas data recordings are made at a SIP endpoint hosted on a professional VoIP system which uses a $\mu$-law codec operating with an 8~kHz sampling rate.  
Condition C3 may hence reflect the application of multiple, unknown intermediate transcodings and network transmissions.
No codec information nor any external metadata was provided in the audio file headers.
The number of trials in the LA progress and evaluation subsets is shown in Table~\ref{tab:All_trial_num}. 
The progress subset contains a modest number of C1-C4 utterances, while the evaluation subset contains the remaining C1-C4 utterances in addition to the full set of C5-C7 utterances. 

The distribution of speakers and spoofing attacks is balanced in each condition, implying that differences in detection performance can be attributed reliably to the variations in encoding and transmission.
While the total number of speakers is the same as that of the ASVspoof 2019 LA evaluation set, the 2021 database contains a substantial number of previously unexposed, new bona fide utterances collected from the same set of speakers in addition to new spoofed utterances generated with the same attack algorithms.

Last, with no new matched training nor development data being provided for the 2021 edition, challenge rules dictate the use for training purposes of \emph{only} ASVspoof 2019 LA training and development subsets which contain only clean data (no similar encoding and transmission). 
Use of the 2019 evaluation subset was strictly forbidden.
The challenge for the ASVspoof 2021 LA task is hence to design spoofing CMs which generalise well to unknown channel variation. 
While the use of external \emph{speech} data was not permitted, the challenge rules allow the use of external \emph{non-speech} resources such as noise samples, impulse responses, and audio compression software for training or data augmentation~\cite{delgado2021asvspoof}.

\subsection{Physical Access (PA)}
\label{sec:data_pa}
Post-challenge analysis of the ASVspoof 2019 PA results showed evidence of over-fitting to simulated attacks;
CM performance for attacks recorded in real physical spaces was generally found to be worse than for simulated attacks~\cite{Nautsch2021}. 
Even so, the use of simulation allows for the generation of plentiful data in an unlimited number of different, simulated physical spaces at little cost and hence remains an attractive alternative to recordings in real physical spaces.
 
\begin{figure}[t!]
    \centering
    \includegraphics[trim=0 560 0 100, clip, width=1.0\columnwidth]{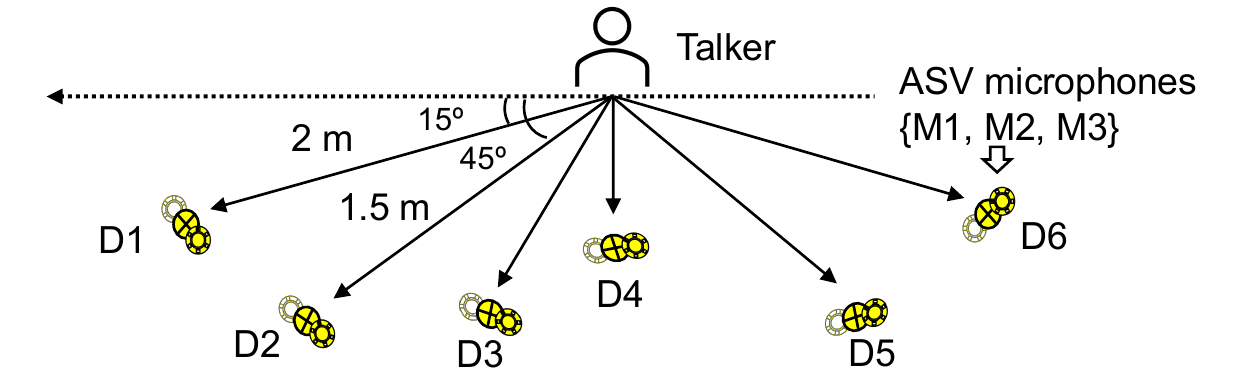}
    \caption{Illustration of talker-to-ASV distance ${D}_{\text{s}}\in\{\text{D1},\text{D2},\cdots, \text{D6}\}$. The full set of ASV microphones $\{\text{M1}, \text{M2}, \text{M3}\}$ are used in all positions. Attacker-to-ASV distance $\mathcal{D}_{\text{s}}'$ can be plotted similarly by replacing the talker with the attacker replay device.  Distances and angles are depicted for D1 and D2 only; those for D3-6 are listed in Table~\ref{tab:PA_conditions}.}
    \label{fig:pa_d_s}
\end{figure}

The ASVspoof 2021 PA task was designed to encourage progress in generalisation, specifically CMs trained and developed using simulated attack data which reliably detect presentation attacks made in real physical spaces.  
The evaluation data hence comprises bona fide speech and replayed recordings both collected in a variety of \textit{real} physical spaces. 
Bona fide data comprises 670 utterances from the VCTK corpus \cite{vctk}\footnote{This is the same set of VCTK utterances used to create the ASVspoof 2019 PA evaluation set~\cite[Fig.~2, \#15]{asvspoof2019database}.} 
which are presented to an ASV system using a high-quality loudspeaker with a reasonably flat frequency response~\cite{PreSonus}. 
Recordings are made in 162 ($=9 \times 3 \times 6$) acoustic environments comprising nine different rooms ${S}_{\text{asv}} \in \{\text{R1}, \text{R2}, \cdots, \text{R9}\}$, three different ASV microphones ${Q}_{\text{asv,m}} \in \{\text{M1}, \text{M2}, \text{M3}\}$, and six different talker-to-ASV distances (and angles) ${D}_{\text{s}}  \in \{\text{D1},\text{D2},\cdots, \text{D6}\}$ listed in Table~\ref{tab:PA_conditions}. The talker-to-ASV distances are illustrated in Fig.~\ref{fig:pa_d_s}.

\begin{table}[t]
    \centering
    \caption{Summary of PA data conditions. \underline{Underlined factors} appear only in the evaluation subset and not in the progress subset. Technical details of microphone and replay devices are available in the supplementary material.}
    \resizebox{\columnwidth}{!}{%
    \setlength{\tabcolsep}{4pt}
    \begin{tabular}{|c|c| c |c|c|}
        \cline{1-2}
         \multirow{2}{*}{$S_{\text{asv}}$ / $S_{\text{a}}$ }    & Room size & \multicolumn{3}{c}{} \\ 
        \cline{4-5}
          & $w\times d \times h (m)$ &  & $D_{\text{s}}$ / $D_{\text{s}}'$ / $D_\text{a}$ & Angle, Dis.(m)  \\ 
        \cline{1-2}\cline{4-5}
          R1 / r1 & $8.0\times8.0\times2.4$ &  &     D1 / d1 \textcolor{white}{/ c1} & $15^{\circ}$,  2.0  \\ 
                                                              R2 / r2 & $6.0\times5.0\times2.3$ &  &                                            D2 / d2 / c2 & $45^{\circ}$,  1.5  \\ 
                                                              \underline{{R3}} / \underline{{r3}} & $6.6\times5.0\times2.4$ &  &                                            D3 / d3 / c3 & $75^{\circ}$,  1.0  \\
                                                              R4 / r4 & $7.5\times7.7\times2.6$ &  &                                            D4 / d4 / c4 & $90^{\circ}$,  0.5  \\
                                                              R5 / r5 & $7.2\times4.0\times2.3$ &  &                                            D5 / d5 \textcolor{white}{/ c5} & $120^{\circ}$, 1.25 \\
                                                              \underline{{R6}} / \underline{{r6}} & $4.5\times6.5\times2.5$ &  &                                            D6 / d6 \textcolor{white}{/ c6} & $150^{\circ}$, 1.75 \\
        \cline{4-5}
                                                              R7 / \underline{{r7}} & $4.5\times2.4\times2.4$ & \multicolumn{3}{l}{\quad ${D}_{\text{s}}$: Talker-to-ASV distance}\\
                                                              R8 / r8 & $7.1\times4.8\times2.5$ & \multicolumn{3}{l}{\quad ${D}'_{\text{s}}$: Attacker-to-ASV distance}  \\ 
                                                              \underline{{R9}} / r9 & $5.9\times4.0\times2.8$ & \multicolumn{3}{l}{\quad ${D}_{\text{a}}$: Attacker-to-talker distance}  \\ 
        \cline{1-2} 
        \multicolumn{2}{l}{${S}_{\text{asv}}$: room for voice presentation}\\
        \multicolumn{2}{l}{${S}_{\text{a}}$: room for replay acquisition}\\[1mm] 
        \cline{1-2} \cline{4-5}
         $Q_{\text{asv,m}}$ / $Q_{\text{a,m}}$ &  Quality &   & $Q_{\text{a, s}}$ & Quality \\ 
        \cline{1-2} \cline{4-5}
          M1 / m1 & Medium & & s2 & Low \\
                                                                     M2 / m2 & High & &   s3 & Medium \\
                                                                     M3 / m3 & Low  & &   \underline{{s4}} & High \\
        \cline{1-2} \cline{4-5}
        \multicolumn{2}{l}{${Q}_{\text{asv,m}}$: ASV microphone} & \multicolumn{3}{l}{\quad ${Q}_{\text{a,s}}$: Attacker replay device}  \\ 
        \multicolumn{2}{l}{${Q}_{\text{a,m}}$: Attacker microphone} \\ 
    \end{tabular}
    }
    \label{tab:PA_conditions}
\end{table}

Replays are made in the same set of rooms according to different attacker factors including the room size ${S}_\text{a} \in \{\text{r1},\text{r2},\cdots, \text{r9}\}$\footnote{We use ${S}_\text{asv}$ and ${S}_\text{a}$ to denote the rooms in which utterances are presented to an ASV microphone and the rooms in which recording of the target speaker's speech are made by the attacker, respectively.}, the attacker microphone device ${Q}_{\text{a,m}} \in \{\text{m1}, \text{m2}, \text{m3}\}$, the attacker replay device ${Q}_{\text{a,s}} \in \{\text{s2},\text{s3},\text{s4}\}$\footnote{\text{s1} denotes the simulated bona fide talker.}, and the attacker-to-talker distance ${D}_{\text{a}} \in \{\text{c2},\text{c3},\text{c4}\}$. 
Attacker replay devices are situated at the same positions as bona fide talkers ${D}'_{\text{s}} \in \{\text{d1},\cdots,\text{d6}\}$\footnote{Different to the notations used in \cite{yamagishi21_asvspoof}, the attacker-to-ASV and the attacker-to-talker distances in this paper are denoted by d* and c*, respectively. c2, c3, and c4 are the same positions as d2, d3, and d4, respectively.}. 
Since the number of exhaustive attack factor combinations is large (1,458 $= 9\times 3 \times 3 \times 3 \times 6$), a non-exhaustive policy was adopted. 
First, the nine rooms are divided into three groups\footnote{$\{$R1/r1, R2/r2, R3/r3$\}$, $\{$R4/r4, R5/r5, R6/r6$\}$, $\{$R7/r7, R8/r8, R9/r9$\}$.} so that each group contains one large, one medium, and one small sized room. 
Replay attacks in each room $S_{\text{asv}}$ are presented to the ASV microphone using a single replay device and speech data recorded in three $S_\text{a}$ rooms in the same room group. 
Recordings are made using the full set of three attacker microphones ${Q}_{\text{a,m}}$, each set at one of the three attacker-to-talker distances ${D}_{\text{a}}$. 
Accordingly, replayed data presented in each room $S_{\text{asv}}$ cover nine different combinations of replay device, attacker room, and attacker microphone. 
The nine combinations are different for each room in which replays are made.

\begin{table}[]
    \centering
    \caption{Summary of DF evaluation conditions. Each of these conditions also includes different vocoder types as sub-conditions. \underline{Underlined factors} only appear in evaluation subset but not in progress subset.}
    \begin{tabular}{|l|c|c|}
        \hline
        Cond. & Compression	& Bitrate    \\ 
        \hline\hline
        DF-C1 & -- & 256~kbps  \\ 
        DF-C2 & Low mp3 & $\sim$80-120~kbps  \\ 
        DF-C3 & High mp3 & $\sim$220-260~kbps  \\
        DF-C4 & Low m4a & $\sim$20-32~kbps \\ 
        DF-C5 & High m4a & $\sim$96-112~kbps \\ 
        \underline{DF-C6} & Low ogg & $\sim$80-96~kbps\\ 
        \underline{DF-C7} & High ogg & $\sim$256-320~kbps \\
        \underline{DF-C8} & mp3$\rightarrow $ m4a & $\sim$80-120~kbps, $\sim$96-112~kbps \\
        \underline{DF-C9} & ogg $\rightarrow$ m4a  &  $\sim$80-96~kbps, $\sim$96-112~kbps\\
        \hline
    \end{tabular}
    \label{tab:DF_conditions}
\end{table}

Recordings and replays are made using VCTK source utterances at a sampling rate of 48~kHz. After collection, all bona fide and spoofed data are downsampled to 16~kHz 
after anti-aliasing --- applying a non-causal Hamming-windowed sinc filter truncated to a length of $2^{14}+1$ samples. This downsampling configuration is the same as that used in collecting the ASVspoof 2019 PA database. Leading and trailing non-speech segments of longer than 0.3 seconds are truncated. The set of trials are then divided into utterance-disjoint progress and evaluation subsets. Both sets include data from the full set of speakers. As a further test of generalisation, a subset of replay trials corresponding to the factors underlined in Table~\ref{tab:PA_conditions} were withheld from the progress subset and reserved exclusively for the evaluation subset. For other factors, 80\% of the bona fide and 70\% of the replay trials are assigned to the evaluation set, with the remaining data constituting the progress subset. The number of trials in each set are listed in Table~\ref{tab:All_trial_num}.

\begin{figure*}[t]
    \centering
    \subfloat[LA]{
    \includegraphics[trim=0 15 0 10, width=0.3\textwidth]{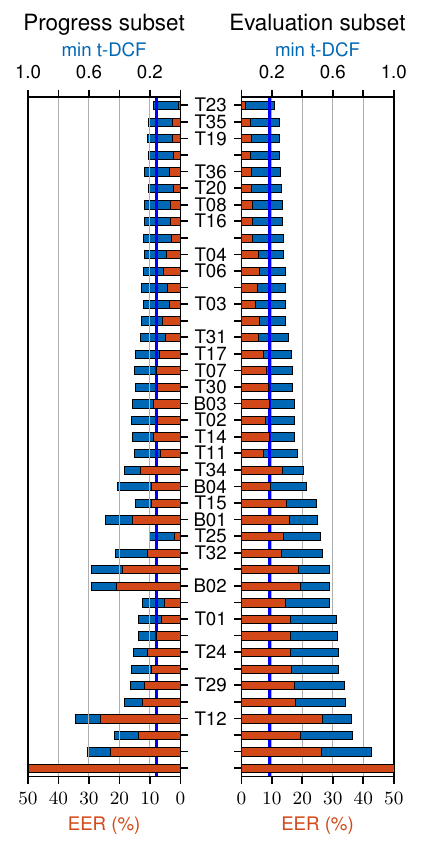}
    }
    \subfloat[PA]{
    \includegraphics[trim=0 15 0 10, width=0.3\textwidth]{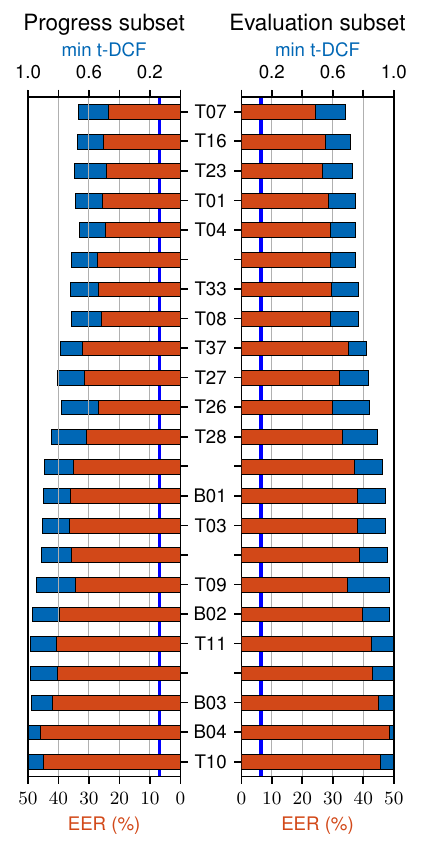}
    }
    \subfloat[DF]{
    \includegraphics[trim=0 15 0 10, width=0.3\textwidth]{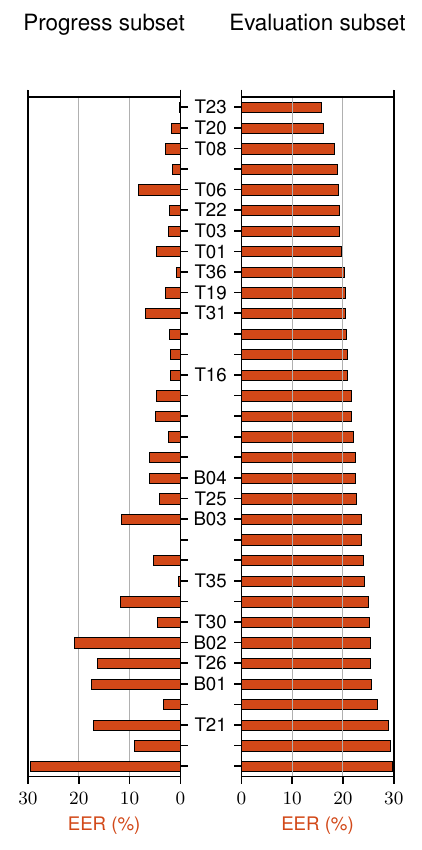}
    }
     \caption{ASVspoof 2021 results for LA, PA, DF tracks. Results shown in terms of \textcolor{colortDCF}{pooled normalised minimum t-DCF} and \textcolor{colorEER}{pooled EER (\%)}. \textcolor{colortDCF}{Blue line} marks ASV floor. Only teams that submitted system description are marked with ID. }
     \vspace{-2mm}
     \label{fig:all_results}
\end{figure*}

\begin{table*}[htbp]
  \footnotesize
  \renewcommand{\arraystretch}{1.1}
  \centering
  \caption{Empirical description of top-5 submissions for each task, and the baseline systems (bottom) \cite{yamagishi21_asvspoof}. Some of those submissions have corresponding published descriptions \cite{top5_system1, top5_system2, top5_system3, top5_system4, top5_system5_t20, top5_system6_t07}. Teams are presented in order according to results of pooled normalized min t-DCF. LFCC: Linear frequency cepstral coefficients \cite{lfcc_mfcc2011}. L-VQT: long-term variable Q transform \cite{l-vqt}. SEnet: Squeeze-and-excitation network \cite{senet}. SFR-CNN: sparse feature reactivation CNN block, from CondenseNetV2 \cite{condensenetv2}.}
  \begin{tabular}{|c|c|c|c|c|c|}
  \hline
    Task & Team ID & Data Augmentation & Acoustic Feature & Classifier & System Fusion \\ \hline
    \multirow{5}{4.25em}{\emph{LA}} & T23 & Trans codec & Raw waveform; Raw spec & LightCNN;ResNet;LSTM & Weighted score average \\ 
    & T35 & Additive noise & LFCC & ResNet & Score average \\ 
    & T19 & RIR;MUSAN & Mel spec & ResNet;SEnet & Weighted score average \\ 
    & T36 & Speed perturbation & Raw Waveform;Mel spec & TDNN;ResNet;MLP & Score average \\
    & T20 & RIR;MUSAN;Media codec & Linear Filter Bank & ResNet;MLP & Score average \\ \hline
    \multirow{5}{4.25em}{\emph{PA}} & T07 & Speed perturbation & Log spec & VAE;GMM & Score average \\
    & T16 & In-house replay & Linear spec;Raw spec;Phase spectrum & SEnet & Score average \\
    & T23 & RIR;MUSAN & Raw spec & ResNet & Weighted score average \\
    & T01 & - & linear spec;Mel spec & LightCNN & Score average \\
    & T04 & RIR;Speed perturbation & Mel spec; CQT spec & TDNN & Score average \\ \hline
    \multirow{5}{4.25em}{\emph{DF}} & T23 & Media codec & Raw Waveform; Raw spec & LightCNN;ResNet;LSTM & Weighted score average \\
    & T20 & RIR;MUSAN;Media codec & Linear filter bank & ResNet;MLP& Score average \\
    & T08 & Trans codec;Media codec & CQT spec & LightCNN & Score average \\
    & T06 & Trans codec;Media codec & Linear spec; Raw spec & SEnet;TDNN;GMM & Weighted score average \\
    & T22 & Media codec & L-VQT spec & SFR-CNN & - \\ \hline \hline
    \multirow{4}{4.25em}{{Baselines \\ LA/PA/DF}} & B01 & - & CQCC & GMM & - \\ 
    & B02 & - & LFCC & GMM & - \\ 
    & B03 & - & LFCC & LCNN-LSTM & - \\ 
    & B04 & - & Raw waveform & RawNet2 & - \\ \hline
  \end{tabular}
\label{tab:results_briefing}
\end{table*}

\subsection{Deepfake (DF)}
Evaluation data for the ASVspoof 2021 DF task is a collection of bona fide and spoofed speech utterances processed with different lossy codecs used typically for media storage. The data is encoded and then decoded to recover uncompressed audio. This process introduces distortions which depend on the codec and its configuration. Generic end-user applications are envisaged, namely we aim to promote solutions for the detection of deepfakes in compressed audio used in television and media hosted on news websites and social media platforms, etc. In contrast to the LA and PA scenarios, the DF task does not involve the use of an ASV system. Accordingly, whereas the \emph{tandem detection cost function} (t-DCF)~\cite{Kinnunen2020-tDCF-fundamentals} is used as the primary metric for LA and PA tasks, that for the DF task is the CM equal error rate (EER).

While the evaluation database originates partly from the ASVspoof 2019 LA evaluation set, there are two additional sets of source data. They are the 2018~\cite{LorenzoTrueba2018-VCC} and 2020~\cite{Yi2020-VCC-challenge} \emph{voice conversion challenge} (VCC) databases. Both are publicly available but neither has been used previously for ASVspoof challenges.
In contrast to LA and PA evaluation data, both of which are derived solely from the VCTK database~\cite{vctk}, the VCC 2018 and 2020 challenge data are derived from DAPS~\cite{Mysore2015-DAPS} and EMIME \cite{Wester2010-EMIME} corpora, respectively. We included all the audio data from the two VCC challenges, including source and target speaker training data and all data submitted by VCC participants. Combined, these two additional source datasets contain bona fide speech collected from a total of 26 additional speakers (12 in VCC 2018, 14 in VCC 2020) and a large number of spoofed utterances generated using VC attack algorithms not used in generating any of the past ASVspoof challenge databases. The aim was to address CM generalisation across different codecs as well as different source databases (domains) and spoofing attacks. Accordingly, the evaluation data consists of various levels of detection difficulty, as described below.

The full DF evaluation database is generated using in excess of 100 different spoofing attack algorithms generated by the large number of teams or individuals who contributed to the ASVspoof 2019 database~\cite{asvspoof2019database}, or by VCC participants. It should still be noted that, since many of the VC approaches share similar voice coders (vocoders), we are most interested in analysing the dependence of CM performance upon broad vocoder categories (rather than individual VC systems). To this end, the set of vocoders are divided into five broad categories: \emph{traditional}, \emph{neural autoregressive}, \emph{neural non-autoregressive}, \emph{waveform concatenation} and \emph{unknown}. The unknown category covers missing or incomplete VC system descriptions.

Audio data from the three sources (ASVspoof 2019, VCC 2018, VCC 2020) was processed with the set of codecs giving the nine evaluation conditions shown in Table~\ref{tab:DF_conditions}. The `no codec' condition C1 corresponds to original audio samples. Conditions C2 and C3 both use an mp3 codec. C4 and C5 use an m4a codec with advanced audio coding,\footnote{\url{https://trac.ffmpeg.org/wiki/Encode/AAC} (referred \today).} whereas both C6 and C7 use an ogg Vorbis codec.\footnote{\url{https://xiph.org/vorbis/} (referred \today).} The differences between conditions with the same codec lie in the use of different, variable bit rate (VBR) configurations, one lower and one higher, as illustrated in the right-most column of Table~\ref{tab:DF_conditions}. Two additional conditions C8 and C9 involve the successive application of two different codecs, one with a lower VBR, the other with a higher VBR. The goal for these two dual-codec conditions is to study whether spoofing artifacts persist in the case of distortion introduced through transcoding. Such transcoding could take place, for instance, when an adversary acquires a sample of compressed spoof speech for a given target speaker from one social media website, and then uploads it to another one that uses a different audio compression technique. We used \texttt{ffmpeg}\footnote{\url{http://ffmpeg.org/}} and \texttt{sox}\footnote{\url{http://sox.sourceforge.net/}} toolkits in creating the DF evaluation data. The number of trials for the DF subset is shown in Table~\ref{tab:All_trial_num}.

\begin{figure*}
 \centering
 \subfloat[LA]{
 \includegraphics[trim=0 15 0 5,  width=.9\textwidth]{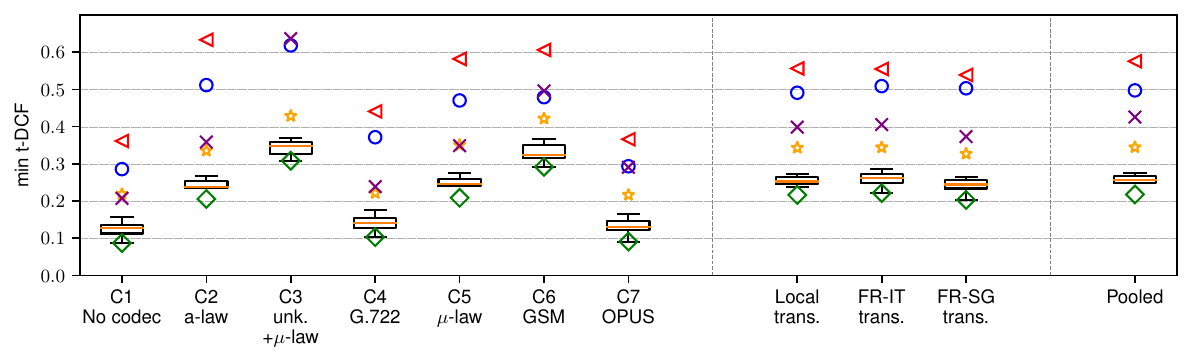}
 \label{fig:LA_Channel_Transmission_decomposed}
 }
 \qquad
\subfloat[PA]{
\includegraphics[trim=0 10 0 0, width=.9\textwidth]{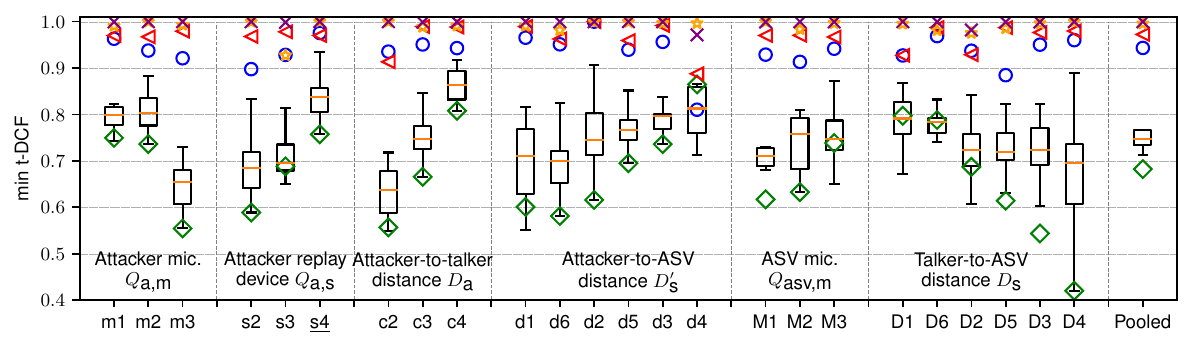}
\label{fig:PA_decomposed}%
}
\qquad
\subfloat[DF]{
 \includegraphics[trim=0 10 0 0, width=.9\textwidth]{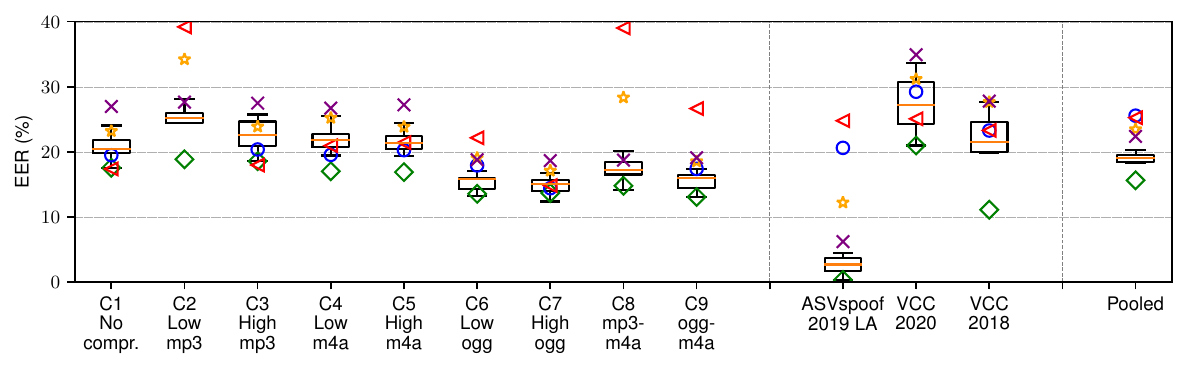}
 \label{fig:DF_subcondition_result_decomposition}}
 \vspace{-1mm}
 \caption{Boxplots of evaluation set min t-DCF or EERs of top-10 submissions decomposed over different factors. Markers are top-1 submission (\textcolor{teal}{$\diamond$}), B01 (\textcolor{blue}{o}), B02 (\textcolor{red}{$\triangleleft$}), B03 (\textcolor{orange}{$\medwhitestar$}), and B04 (\textcolor{purple}{$\times$}).}
\end{figure*}

\begin{figure}[t!]
    \centering
    \includegraphics[trim=0 15 0 0, width=0.9\columnwidth]{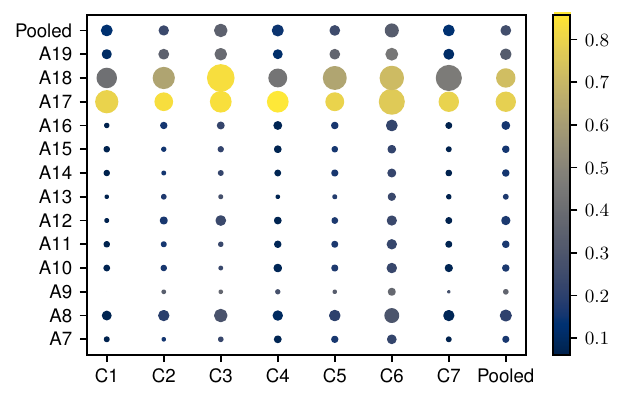}
    \caption{Statistics on t-DCF values of top-10 LA submissions in sub-conditions. Dot size is decided by inter-quartile range of t-DCF distribution, and color is decided by its median.}
    \label{fig:la_subcondition_analysis}
\end{figure}

\begin{figure}[t!]
    \centering
    \includegraphics[trim=0 10 0 0, width=0.85\columnwidth]{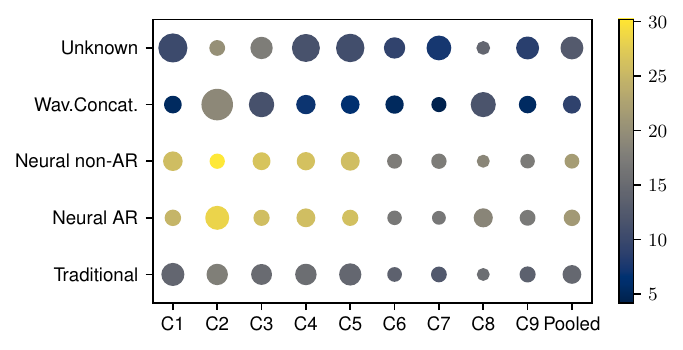}
    \caption{Break-down of the EER in the VCC 2018/2020 subset of the DF track, according to the codecs and the type of vocoder used by the VC system. 
    }
    \label{fig:cross-factor-EER-DF}
\end{figure}

\section{ASVspoof 2021 Challenge Results}
\label{sec:results}
A summary of results for all three tasks is shown in Fig.~\ref{fig:all_results}. Baseline systems (B0$n$) and submissions (T$nn$)\footnote{Submissions with no identifier correspond to teams that failed to submit a valid system description. Due to ASVspoof challenge policy, which allows for anonymous participation, neither the team nor individuals' names connected to a given team identifier can be revealed.} are ranked according to their performance on the evaluation subset and primary metric. For all three tasks, CM performance is illustrated in terms of the EER in red bars (bottom axes), which is the primary metric only for the new DF task. For LA and PA tasks, the primary metric is the min t-DCF \cite{Kinnunen2020-tDCF-fundamentals}. These results are shown in blue bars (top axes). The \emph{ASV floor}~\cite[Eq. (11) and Section V]{Kinnunen2020-tDCF-fundamentals}, which corresponds to the t-DCF in the case of a perfect CM but imperfect ASV system, is illustrated with a vertical blue line. The gap between blue bars and blue lines give an indication of CM performance. Table~\ref{tab:results_briefing} shows a description of the best performing submissions for each task, along with the one for the baseline systems.

\subsection{Logical Access (LA)}

\subsubsection{Full challenge results}
Results for the LA task are shown in Fig.~\ref{fig:all_results}a. Many CMs outperform the best baseline B03, some by a large margin. The top systems show t-DCFs close to the ASV floor, indicating CMs with low error rates. For the majority of systems the gap in performance for progress and evaluation subsets is modest. The over-fitting to the progress subset seen for other systems is likely caused by differences between the known (C1-C4) and unknown (C5-C7) conditions, the latter of which appear only in the evaluation subset.

\subsubsection{Influential data factors}
The left-most pane of Fig.~\ref{fig:LA_Channel_Transmission_decomposed} shows the distribution in t-DCF for the top-10 systems for each of the seven conditions shown in Table~\ref{tab:LA_conditions}. For reference, the distribution for pooled conditions is shown in the right-most pane. The distribution in t-DCF for wideband conditions (no codec, G.722, OPUS) is lower than for narrowband conditions (a-law, PSTN, u-law and GSM), indicating the general importance of information at higher frequencies to CM performance. Among the narrowband conditions, lower bit rates and uncontrolled transmission (GSM and unk.\ + $\mu$-law, respectively -- see Table~\ref{tab:LA_conditions}) lead to worse performance.

The middle pane of Fig.~\ref{fig:LA_Channel_Transmission_decomposed} shows the distribution in t-DCF for the three transmission routes detailed in Section~\ref{sec:LAdb}, namely routes across a LAN and from France to Italy (FR-IT) and from France to Singapore (FR-SG).  The similarity in t-DCF for all three shows that transmission routes have little impact upon CM performance; they are not dissimilar to the distribution for pooled results. This would suggest that future challenge editions could use simpler LAN routes only.

Fig.~\ref{fig:la_subcondition_analysis} depicts a condition/attack analysis, with attack algorithms on the vertical axis and evaluation conditions on the horizontal axis. The color of each circle encodes the median min t-DCF for the top-10 systems, while the radius represents the inter-quartile range. For some attacks and some conditions the min t-DCF is consistently high. For attack A18, and to a lesser extent also A17, there is greater variation across conditions, with the median t-DCF ranging from approximately 0.4 for condition C1 (no encoding) to over 0.8 for C3 (unk.\ + $\mu$-law) and approximately 0.7 for C6 (GSM). These observations show that A18, an attack based on a non-parallel VC system (details in Section 3.1, \cite{asvspoof2019database}) which is already difficult to detect even in the absence of encoding, becomes disproportionately more difficult to detect after low bandwidth encoding.

\subsubsection{Top-performing systems}
A summary of the top-5 systems is presented to the top of Table~\ref{tab:results_briefing}. All use some form of data augmentation, though no common form can be identified. Most are ensemble systems, and most operate upon short-term spectral features or raw waveforms. Most use a ResNet classifier or variant, with other types of convolutional networks also being popular. Fusion strategies include weighted averaging, using either uniformly or empirically set weights.

\subsection{Physical Access (PA)}

\subsubsection{Full challenge results}
A summary of results for the PA task is shown in Fig.~\ref{fig:all_results}b. While almost half of the systems outperform the best B01 baseline, the performance of all systems is substantially worse than the ASV floor of 0.12. This would suggest that the PA task is more challenging than the LA task. The difficulty might be attributed to the difference between simulated replay attacks (data used for training and development)~\cite[Section 2.2.3]{asvspoof2019database} and replay attacks recorded in real physical spaces (evaluation data), e.g., differences in room acoustics and noise conditions. We also observe only modest gaps between performance for progress and evaluation subsets indicating that, while error rates remain high, system performance is stable across different evaluation conditions (rooms and devices, see Section~\ref{sec:data_pa}).

\subsubsection{Influential data factors}
Fig.~\ref{fig:PA_decomposed} shows min t-DCF results pooled over the top-10 and baseline systems for a selection of attacker and environment factors. The rooms used for replay acquisition and voice presentation are not included in the analysis because we observe no substantial correlation between the room size and the min t-DCF values. However, we observe expected impacts related to the quality of the attacker microphone. For example, higher-quality attacker microphones m1 and m2 lead to higher min t-DCFs than the lower-quality microphone m3. This is to be expected because spoofed data acquired using a higher-quality microphone  $Q_{\text{a,m}}$ introduces less distortion and are hence more challenging to detect. 

The min t-DCF is also higher for a better quality attacker replay device $Q_{\text{a,s}}=\text{s3}$ or $\text{s4}$. Similarly, shorter attacker-to-talker distances $D_\text{a}$ also lead to worse performance. At the closest position $D_\text{a}=\text{c4}$, min t-DCF values are the highest.

The attacker-to-ASV distance $D'_{\text{s}}$ is observed to be inversely correlated with the min t-DCT. However, as expected, if the talker is closer to the ASV microphone (i.e., moving from $D_\text{s} = \text{D1}$ to D4), then the min t-DCF decreases.

\subsubsection{Top-performing systems}
A summary of the top-5 PA systems is presented in the middle of Table~\ref{tab:results_briefing}. Similar to the LA task, most use data augmentation, once again with various different approaches, and all are ensemble systems. Different to observations for the LA task, there is little variation among front-end features used for the PA task, while there is diversity among the adopted classifiers. The top-1 system uses both frame-level and temporal-level features and the parallel combination of variational autoencoder (VAE) and Gaussian mixture model (GMM) classifiers. Like for the LA task, all the top-5 systems use score averaging.

\subsection{Deepfake (DF)}

\subsubsection{Full challenge results}
A summary of results for the DF task is shown in Fig.~\ref{fig:all_results}c. The main observation is the striking difference in results for progress and evaluation subsets. While 23 (out of 33) systems have EERs of less than 10\% for the progress subset, and while the best performing system even has an EER of less than 1\%, \emph{all} have EERs exceeding 15\% for the evaluation set. This indicates the high level of difficulty indicated by the underlined conditions and their associated compression algorithms in Table \ref{tab:DF_conditions}. Despite high EERs, however, 18 systems outperformed the best B04 baseline system.

\subsubsection{Influential data factors}
The left-most pane of Fig.~\ref{fig:DF_subcondition_result_decomposition} shows the distribution in EER performance for each compression method, with the distribution for pooled conditions being shown in the right-most pane. For mp3 compression, a higher VBR (C3) gives a lower median EER than for a lower VBR (C2). For m4a and ogg codecs, there is no similar effect between the low and high bit rates for either codec. EERs for the ogg codec are lower than those for both mp3 and m4a codecs, no matter what the VBR. Interestingly, EERs for the two double compression methods are only modestly higher than those for the single ogg compression, but below those for either mp3 or m4a compression, indicating some resilience to transcoding.

The middle pane of Fig.~\ref{fig:DF_subcondition_result_decomposition} shows the distribution in EER for each source database. EERs are substantially higher for VCC datasets than for the ASVspoof 2019 LA database. This is likely due to the use of the latter in the progress set only and is hence a consequential over-fitting, or lack of generalisation to mis-matched VCC source data. Further analysis of this observation is provided in Section~\ref{subsec:DF-high-EER}.

Finally, Fig.~\ref{fig:cross-factor-EER-DF} shows a breakdown in the EER for pooled VCC 2018 and 2020 source data according to the type of vocoder and codec. The results indicate that, for any given codec, both types of neural vocoders yield higher EERs than more traditional vocoders or waveform concatenation.

\subsubsection{Top-performing systems} 
A summary of the top-5 DF submissions is illustrated to the bottom of Table~\ref{tab:results_briefing}. With one exception, all submissions are ensemble systems. All use data augmentation.
While there are different approaches, all use some form of media codec augmentation.\footnote{We use \emph{media codec} to refer to generic codecs implemented in various software packages. The challenge participants used mp3, m4a, ogg, opus, mp4, aac, alac, vorbis, wma and sbc, along with other open-source toolkits for media codec augmentation.} Acoustic features are diverse. 
A range of different classifiers are also used, including mostly convolutional neural networks, but with simpler MLP and GMM classifiers also featuring among the top-5 submissions.
Once again, all top-5 systems use score averaging for fusion.

\section{Hidden Subsets and Additional Analyses}\label{sec:Hidden-Tracks}
We now turn attention to results for the hidden subsets first mentioned in Section~\ref{sec:data_pa} before presenting other analyses. The purpose of hidden subsets is to assess the dependence of CM performance upon certain data characteristics that were withheld from challenge participants. They were unaware of their inclusion and treated hidden subsets in the same manner as evaluation data. Hidden subsets were not used in deriving challenge results or rankings.

\subsection{The role of non-speech}
The databases for all three tasks contain a hidden subset of utterances from which non-speech intervals are automatically removed.
The motivation is to assess performance when spoofed \emph{speech} detectors are constrained to operate only upon \emph{speech} segments; 
recent reports (e.g.~\cite{Chettri2020-dataset-artefacts, zhang21da_interspeech, muller21_asvspoof}) suggest that some CMs might rely upon information within non-speech segments. 
While non-speech information might serve as legitimate cues for spoofing detection, adversaries could easily remove non-speech segments to gain an advantage.
Alternatively, cues within non-speech intervals might correspond to unintended database design artifacts.
It is hence of interest to determine any differences in CM performance when they are restricted to using information from speech segments only. 

\begin{figure*}
\centering
\subfloat[LA]{
\includegraphics[trim=0 5 0 0, width=0.29\textwidth]{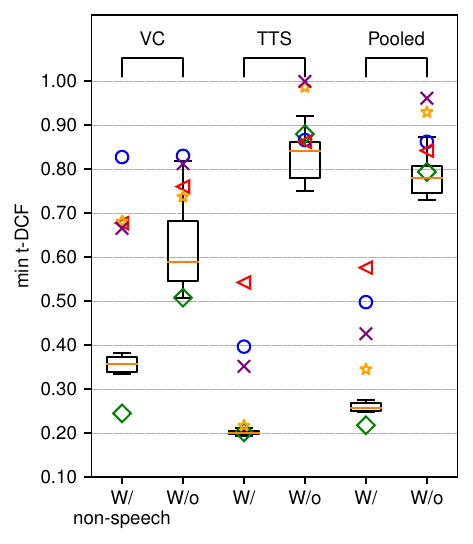}
}
\subfloat[PA]{
\includegraphics[trim=0 5 0 0,width=0.29\textwidth]{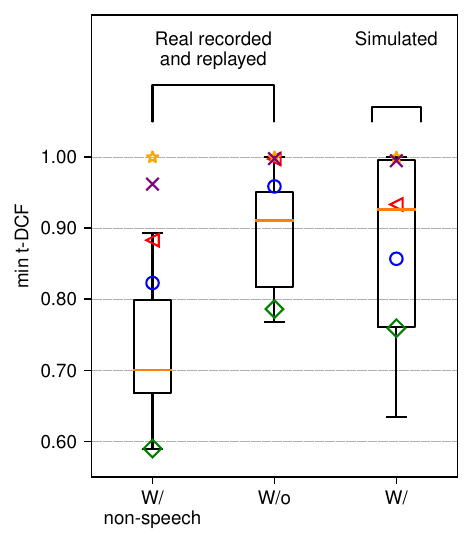}
}
\subfloat[DF]{
\includegraphics[trim=0 5 0 0,width=0.29\textwidth]{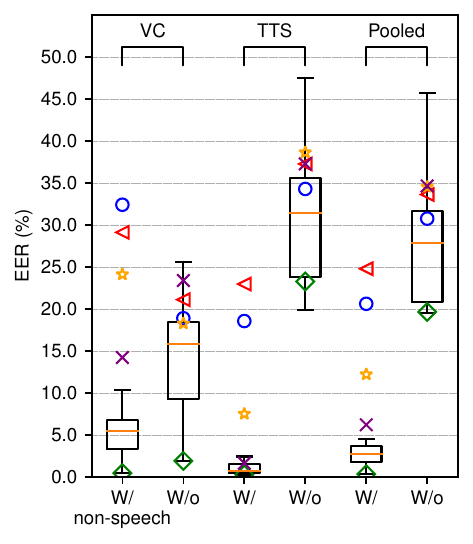}
}
\caption{Comparison of top-10 CMs performance on evaluation subset (with non-speech) and hidden subset (w/o non-speech). For LA and DF, min t-DCFs over VC- and TTS-based spoofing attacks are plotted separately. Markers are top-1 system (\textcolor{teal}{$\diamond$}), B01 (\textcolor{blue}{o}), B02 (\textcolor{red}{$\triangleleft$}), B03 (\textcolor{orange}{$\medwhitestar$}), and B04 (\textcolor{purple}{$\times$}).}
\label{fig:hidden_track_decomposed}%
\end{figure*}

Bona fide and spoofed utterances are selected at random from evaluation subsets giving the number of utterances shown in Table~\ref{tab:All_trial_num} for LA, PA and DF tasks.
An energy-based voice activity detector (VAD) with the recommended threshold settings~\cite[Section 5.1]{kinnunen2010overview} is then used to remove segments labeled as non-speech in addition to any speech segments of less than 50~ms duration. For the LA task, the hidden subset contains C1-C7 utterances that are all treated by a VAD to remove non-speech segments. So as to focus upon speech/non-speech effects only, rather than variations in room characteristics and replay configurations etc., the PA hidden subset contains data from the D4 talker-to-ASV distance only (90$^{\circ}$, 0.5~m from the ASV microphone as per Table~\ref{tab:DF_conditions}). The hidden subset for the DF task comprises utterances for all conditions C1-9, but ASVspoof 2019 LA source data only.

Contrasting min t-DCF/EER results for evaluation data (w/non-speech) and hidden subset (w/o non-speech segments) are plotted in Fig.~\ref{fig:hidden_track_decomposed} for the LA task (a), the PA task (b) and DF task (c).
Boxplots for the top-10 LA systems are shown separately for VC-based (left), TTS-based (middle) and pooled (right) attacks.\footnote{Following the descriptions in \cite{asvspoof2019database}, A17, A18, and A19 are grouped as VC-based attacks.  All others are grouped as TTS-based attacks.} 
While the contrast is greatest for TTS attacks, min t-DCFs calculated from utterances containing non-speech segments (w/non-speech) are substantially lower than those calculated from utterances without non-speech segments (w/o).
LA source data contain long non-speech segments at the start and/or end of each utterance. While VC spoofing attack algorithms reproduce these characteristics, TTS algorithms produce utterances without such long non-speech segments. Their length may hence serve as a cue to distinguish between bona fide and spoofed utterances, accounting for why their absence leads to degraded performance. The reliance upon such cues may not lead to reliable detection in the wild. Observations for the DF task shown in Fig.~\ref{fig:hidden_track_decomposed}c are much the same as for the LA task.
Without the use of non-speech segments, performance is substantially degraded, more so for TTS attacks than VC attacks.

Boxplots for the top-10 PA systems are shown in Fig.~\ref{fig:hidden_track_decomposed}b for replay attacks recorded in real physical spaces, with (evaluation subset) and without (hidden subset) non-speech segments (two left-most boxes). The length of non-speech intervals is not expected to have an impact since  recording and replaying do not alter utterance duration. Nonetheless, without non-speech segments, performance is notably worse. Any asymmetry in the \emph{length} of non-speech segments in bona fide and spoofed speech should not serve as a detection cue. The \emph{contents} of non-speech segments, however, is a different issue. Non-speech segments, while being information that is easily removed by the adversary, are an innate characteristics of recorded speech which might still provide legitimate spoofing detection cues.

\subsection{Real v/s simulated replay attack}
Another hidden subset in the PA database contains \emph{simulated} replayed data.\footnote{In the official evaluation package, the simulated replayed data and the data without non-speech are referred to as hidden track 1 and 2, respectively.}
The motivation is to determine whether reliable estimates of performance can also be derived using evaluation data that is also simulated instead of being recorded in real physical spaces.
Simulated bonafide and spoofed data are created using the same tool as that used in generating the ASVspoof 2019 PA database~\cite[Section 4.3]{asvspoof2019database}. 
Room impulse responses and background noises are extracted from swept-sine signals recorded in the same rooms and with the same set of microphones. For the same reasons outlined above,
data is simulated for the D4 talker-to-ASV and the d4 attacker-to-ASV distance only.

Results shown in the right-most pane of Fig.~\ref{fig:hidden_track_decomposed}b depict the distribution in min t-DCFs for the top-10 PA systems for the hidden, simulated subset which contains non-speech intervals. 
The median min t-DCF for the simulated subset is much higher than that for the evaluation subset. 
Further analysis shows that four out of the top-10 PA systems' min t-DCFs for the simulated subset are higher than 0.99. 
The gap between the results on the hidden and evaluation sets suggests that the simulated subset looses the artefacts in the real recording and replaying environments that are helpful to discriminate bona fide and spoofed data. Analysis in the supplementary material (Sec. III.C) found that the hidden track data involving various simulated rooms and devices is equally challenging to most of the CMs. This is different from the observations from Fig.~\ref{fig:PA_decomposed} on the real recorded and replayed data. The results hence caution against the use of simulated evaluation data to estimate CM performance in real physical environments if it cannot faithfully reflect the room acoustics and frequency response characteristics of the devices. 

\begin{figure*}
    \centering
    \includegraphics[scale=0.33]{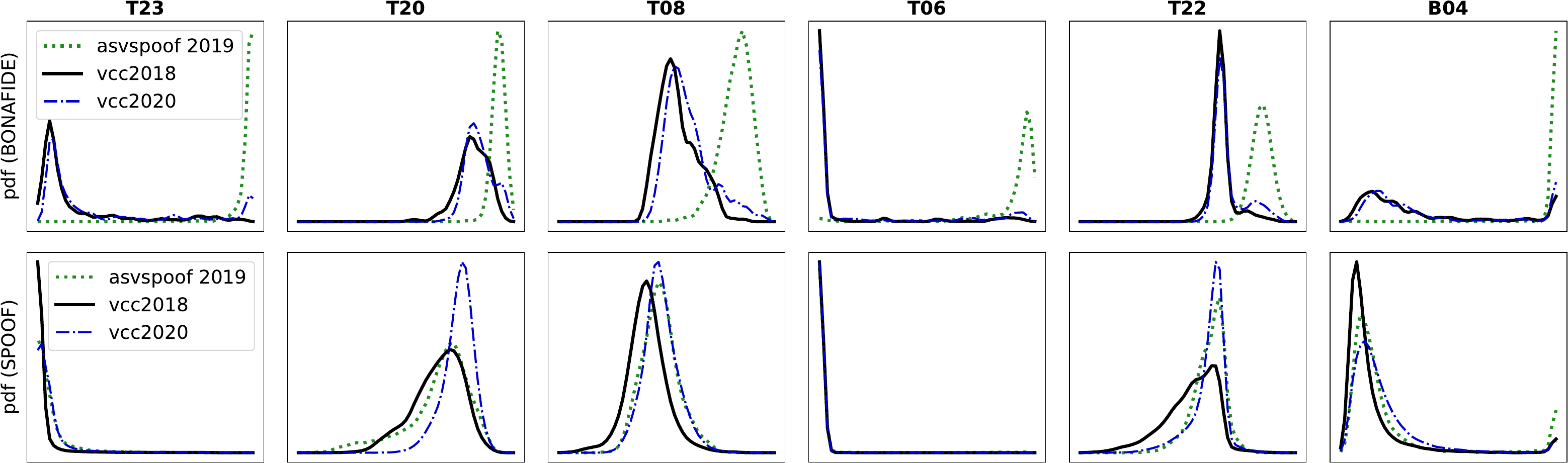}
    \caption{Score distributions of bonafide (top) and spoof (bottom) for  top-5 DF subsystem and baseline B04, broken down according to three data sources used.}
    \label{fig:DF-score-plot}
\end{figure*}

\begin{table}[!t]
    \centering
    \caption{EERs (\%) for the DF track with all compressors pooled, but bonafide/spoof trials selected from different combinations of source datasets.}
    \setlength{\tabcolsep}{5pt}
    \begin{tabular}{|cc||cccccc|}
         \hline
         Bonafide & Spoof   & & & & & &\\
         trials   & trials & T23   & T20   & T08   & T06   & T22   & B04\\
         \hline\hline
         All    & All      & 15.63 & 16.04	& 18.29	& 19.01	& 19.22 & 22.38\\
         VCC18  & All      & 23.39 & 27.64	& 31.03 & 30.87 & 29.81 & 34.44\\
         VCC20  & All      & 16.98 & 19.77 & 21.55 & 22.88 & 25.00 & 31.48\\
         ASV19  & ASV19    & 0.33 &	 2.41 &	3.75 & 8.82 & 2.76 & 6.21\\
         ASV19  & VCC18+20 & 0.60 & 1.98 &	2.78    & 5.83  & 1.93  & 3.09\\
         ASV19  & All      & 0.55  & 2.05     & 2.92  & 6.32  & 2.11  & 3.56\\
         \hline
    \end{tabular}
    \label{tab:DF-trials-from-different-source-subsets}
\end{table}

\begin{table}[!t]
    \centering
    \caption{Similar to Table~\ref{tab:DF-trials-from-different-source-subsets}, but using the proportion of detected nonspeech frames by VAD as spoofing detection score instead of output via CM system. Only trials of C1 were considered.}
    \setlength{\tabcolsep}{5pt}
    \begin{tabular}{|ll||c|}
         \hline
         Positive set & Negative set & EER\\
         \hline\hline
         All bona    & All spoof      & 37.59 \\
         VCC18 bona  & All spoof     & 44.17 \\
         VCC20 bona  & All spoof     & 51.14 \\
         ASV19 bona  & ASV19 spoof    & 24.98 \\
         ASV19 bona  & VCC18+20 spoof & 9.19 \\
         ASV19 bona  & All spoof      & 10.36 \\
         \hline
         ASV19 bona  & VCC18 bona   & 11.90  \\
         ASV19 bona  & VCC20 bona   & 14.00  \\
         VCC18 bona  & VCC20 bona   &  43.14 \\
         \hline
    \end{tabular}
    \label{tab:DF-silence-proportion-as-score-from-different-source-subsets}
\end{table}

\subsection{Performance gap for DF progress and evaluation subsets}
\label{subsec:DF-high-EER}

Fig.~\ref{fig:all_results} shows a substantial gap in performance for progress and evaluation subsets for the DF task. 
As evident from results shown in Fig.~\ref{fig:DF_subcondition_result_decomposition}, the gap relates in part to the inclusion of two previously unexposed source corpora. 
To explore the cause for this gap, we examined the class- and data source-conditional CM score distributions for the top-5 submissions.
Results are shown in Fig.~\ref{fig:DF-score-plot} along with distributions for the best B04 baseline. 
While similar findings were observed for compressed data conditions, the plots shown correspond to non-compressed data (condition C1 in Table~\ref{tab:DF_conditions}).

While, for any given system, score distributions for the spoofed class (bottom row) are reasonably well aligned, there are significant differences in the distributions for the bona fide class (top row).
Scores for the two VCC source corpora are consistently lower than those for the ASVspoof 2019 LA corpora.  
This shift leads to greater overlap in the distributions for bona fide and spoofed classes.
The use of only ASVspoof 2019 LA data for training leads to over-fitting and models which generalise poorly to VCC data in the evaluation subset, greater confusion between bona fide and spoofed trials, and hence degraded detection performance.

As a further numerical quantification, Table \ref{tab:DF-trials-from-different-source-subsets} displays the EERs with either bonafide or spoof trials constrained to specific subsets. 
We find that whenever either one of the two VCC subsets is included on bonafide side (the first three rows), the EERs are very high ($\sim 15\% \dots 34\%$). 
In stark contrast, however, when both VCC'18 and VCC'20 trials are \emph{excluded} from bonafide side (the last three rows), the EERs of the six CMs are substantially lower ($< 1\%$ for T23) -- regardless of whether the \emph{spoof} trials include VCC'18 or VCC'20 data. 

These findings, and given the performance mismatch between evaluation and hidden track which excludes silences, motivated a deeper exploration of the possible role of silence in the VCC bonafide score distribution shift. To this end, for each audio segment from C1 (pooled from the progress and evaluation subsets), the \emph{proportion of detected nonspeech}, defined as the number of detected nonspeech frames detected by VAD divided by the total number of frames, was calculated and used as spoofing detection score. EER computed from these scores reflects the degree of similarity of the corresponding positive and negative trial sets in terms of nonspeech proportion. These EERs, for different combination of positive and negative trials, are shown in Table~\ref{tab:DF-silence-proportion-as-score-from-different-source-subsets}.
The trends are similar to those shown in Table~\ref{tab:DF-trials-from-different-source-subsets}; EERs are substantially lower when ASVspoof 2019 bonafide data is used as positive set, particularly when the negative set contains VCC 2018 and VCC 2020 spoof data. The last three rows show an alternative experiment were both positive and negative sets consists of bona fide data only. EER values between 11\% and 14\% when comparing ASVspoof 2019 and VCC indicate that the nonspeech proportion distribution for ASVspoof and VCC bonafide data are reasonably well separated. However, this does not occur when comparing VCC18 against VCC20 bonafide sets (last row). It would not be a surprise that submitted CMs learned the bonafide class based (at least, partially) on the implicit different silence distribution of the ASVspoof bonafide data (the only data source included in the training set). This in turn makes them fail when an unseen bonafide dataset does not follow the same silence distribution.

\section{Post-Challenge Studies and Related Work}
Each edition of ASVspoof addresses new challenges but also leads to the identification of new challenges and concerns which have stimulated a number of post-challenge studies. The 2021 edition is no exception.
A selection of such studies is presented below, with a summary of related results shown in Table~\ref{tab:post_challenge_results}.
\footnote{The selection is based on searches of ``ASVspoof 2021" and ``Deepfake" in the ISCA archive and IEEExplore.} 
They pertain to LA and DF tasks and (1) the shift from the use of acoustic features toward \textbf{end-to-end} (E2E) architectures with raw waveform inputs and (2) the popularity of \textbf{data augmentation} (DA). 

Originating from~\cite{rawnet2_spoofing2021} and based on RawNet~\cite{rawnet2} and graph attention networks, AASIST~\cite{aasist2022} is representative of progress in the use of E2E architectures.
The RawNet in AASIST has further been replaced by various \emph{self-supervised learning} (SSL) front-ends, including wav2vec \cite{wav2vec_antispoofing2022, ssl_frontends_asvspoof2021, wav2vec_cnn_encoder_add2022}.
While the use of SSL brings substantial improvements in performance, the use of models pre-trained on external data is not compliant with the ASVspoof training protocol. 
A study of different SSL frontends and related techniques can be found in~\cite{ssl_frontends_asvspoof2021}. 
While conventional acoustic features remain popular~\cite{dataaug_study_asvspoof2021}, E2E architectures and SSL-based frontends have led to substantial improvements in performance for the LA task, even for single systems (not ensembles), with best t-DCF being 0.2066 (best EER 0.82\%) for the ASVspoof 2021 LA database~\cite{wav2vec_antispoofing2022}. Such result comes from the use of SSL with data augmentation. Post-challenge studies that employ ASVspoof 2021 PA data and work on replay attacks are rarely seen.

An investigation of DA for the E2E solution which operates directly upon raw waveform inputs, named \emph{RawBoost}, was reported in~\cite{rawboost2022}. The approach is based upon the combination of linear and non-linear filtering in addition to signal-dependent and signal-independent additive noise. 
Investigation on DA techniques applied on anti-spoofing can be found in~\cite{dataaug_study_asvspoof2021,wav2vec_cnn_encoder_add2022,fathan22}. In these studies a number of approaches to DA have been used. Compression, codec and channel effect, bandwidth difference augmentation, as well as time- frequency-masking~\cite{specaugment}.
All these studies indicated that the robustness of spoofed audio detection is substantially improved when DA techniques are employed.
We accordingly provide additional studies with controlled experiments on various types of features, backends, training loss functions, and DA methods using one of the baseline systems for LA and DF in the supplementary material. The experiments follow the ones described in \cite{ssl_frontends_asvspoof2022}.

Finally, other anti-spoofing / deepfake detection challenges and databases have emerged since 2021. The first \emph{audio deep synthesis detection} (ADD) challenge~\cite{add2022} was launched to evaluate the threat of low-quality and partially-spoofed audio~\cite{partialspoof} in order to address real-time and more challenging scenarios. The summary of the results indicate the fragility of synthesis detection model against lower quality and unseen fake audio, especially when partially-spoofed audio is presented.
The spoofing-aware speaker verification (SASV) challenge~\cite{Jung2022SASV}, which exploits the same ASVspoof 2019 LA source database as the ASVspoof 2021 LA evaluation, provides protocols for the design of not just CM solutions, as in the case of ASVspoof, but also ASV systems and alternative architectures for their combination.
Three different metrics support the assessment of spoofing robustness, automatic speaker verification and combined, spoofing aware speaker verification. 
A summary of SASV results~\cite{Jung2022SASV} anticipates the significance of improving the reliability of both ASV and CM sub-systems, including the joint optimization schemes. 

\begin{table}[!t]
    \centering
    \caption{Selected post-challenge results from published papers. 
    Top-1 submission in main challenge and two best baselines are added for reference.}
    \begin{tabular}{|c|r||c|c|c|c|c||c|}
    \cline{1-4}\cline{6-8}
    & &     min t-DCF   &  EER (\%) & & & & EER (\%) \\
    \cline{1-4}\cline{6-8}
    \multirow{7}{*}{LA} & \cite{wav2vec_antispoofing2022} &   0.2066 &  0.82 & & \multirow{7}{*}{DF}  & \cite{wav2vec_antispoofing2022} & 2.85 \\ 
    & T23 & 0.2176 &  1.32  & &  &\cite{wav2vec_cnn_encoder_add2022} & 4.98\\ 
    & \cite{wav2vec_cnn_encoder_add2022} & 0.2780 & 3.54  & & & \cite{ssl_frontends_asvspoof2021} & 6.18 \\
    & \cite{dataaug_study_asvspoof2021} & 0.2882 & 4.66 &  & & \cite{dataaug_study_asvspoof2021} & 14.27 \\
    & B03 & 0.3445 & 9.26 &  & & T23 & 15.64  \\
    & \cite{ssl_frontends_asvspoof2021} & 0.3590 & 7.18 &  & & B04  & 22.38 \\
    & B04 & 0.4257 & 9.50 & &  & B03  & 23.48 \\
    \cline{1-4}\cline{6-8}
    \end{tabular}
    \label{tab:post_challenge_results}
\end{table}

\section{Limitations and future directions}
While we are hopeful that ASVspoof has made substantial contributions to open data and reproducible research in anti-spoofing, like any evaluation, it is not without limitations. 
Some of the most relevant issues that are fundamental to the future ASVspoof road-map are discussed in the following.

\textbf{Realistic audio data} -- ASVspoof 2021 made important steps towards evaluation with data that is more representative of realistic conditions.  Other steps remain to be taken, especially the consideration of background noise for the LA task, the collection of recordings from human talkers as opposed to loudspeakers for the PA task, and the use of real social media for the DF task. 

\textbf{Role of non-speech} -- The results for hidden data sets confirm that non-speech intervals can play a role in the detection of spoofed speech, particularly for LA and DF tasks. 
While non-speech intervals are an innate characteristic of natural speech, and while their contents can help to distinguish between bona fide and spoofed speech, their length is a database characteristic that should not serve as a cue for detection. 
It is of interest to investigate the generation or conversion of non-speech regions in the case of TTS and VC attacks as well as the difference between cues used for detection in non-speech and speech intervals.

\textbf{Training data policy} -- The use of a strict training protocol ensures the meaningful comparability of results generated by different teams and is fundamental to most benchmarking evaluations. 
Even so, post-evaluation results show the benefits of a relaxed training policy which allows the use of larger, more complex models trained using external data, e.g. via semi-supervised learning.  
To allow for their further exploration in a benchmarking evaluation setting, future editions of ASVspoof could hence allow a dual training policy, one fixed, the other relaxed.

\textbf{Diversity of attacks} -- The spoofing attacks in the 2021 LA and DF databases were generated with TTS and VC algorithms that were at the state of the art prior to 2020.
TTS and VC technology has advanced rapidly in recent years, especially in terms of requiring far less target speaker training data~\cite{cooper2020zero, huang2022meta, pmlr-v162-casanova22a}.
The next ASVspoof edition calls for a renewed spoofed data collection and generation effort and the exploration of vulnerabilities to the latest attack techniques.

\textbf{Diversity in audio data} -- ASVspoof data are exclusively English-language, read-speech sourced from the VCTK corpus~\cite{vctk}. The inclusion of two unexposed datasets in the DF database exposed a lack of CM generalization. The cause is due mostly to differences in the characteristics of \emph{bona fide} speech and score distributions. This may suggest over-fitting to speakers in the training datasets. In the future, ASVspoof should increase diversity not only in spoofing attacks, but also in bona fide source data, including collection in different recording environments, languages and speaking styles, among other factors.

\textbf{Extensible evaluation} -- The two ASV-related tasks, LA and PA, used a fixed ASV system, with participants developing only the spoofing CM. 
The literature shows an increasing interest in the joint optimisation of both CMs and ASV systems and alternative combination architectures. 
The spoofing-aware speaker verification (SASV) challenge~\cite{Jung2022SASV} was launched to support research in this direction.  
SASV challenge results~\cite{Jung2022SASV} have shown the need for larger labeled datasets for the learning of speaker embedding extractors and end-to-end solutions. 
ASVspoof and SASV will likely merge for a common evaluation in the future.
This will require not only a huge data collection effort but further thought to evaluation metrics; ASVspoof uses an ASV-constrained t-DCF, whereas the SASV challenge uses a set of three different EER-based metrics.

\textbf{Integration of multiple scenarios} --
There is no reason why replay attacks cannot be made in LA scenarios (variable microphone, telephony applications), nor why TTS and VC attacks cannot be replayed in PA scenarios (fixed microphone, access control applications).
Their exploration will increase the complexity of both protocol design and data collection, but remains an important future direction for ASVspoof.

\textbf{Partially spoofed audio} --
ASVspoof has considered only the scenario in which utterances are either bona fide or spoofed in \emph{full}.
\emph{Partial} spoofs, where only a potentially short interval or perhaps a single word is substituted or manipulated, are known to pose a greater detection challenge~\cite{partialspoof}.
The partially spoofed scenario is perhaps most relevant to the DF task in which there is no ASV system.
This is because partially-spoofed utterances of the shortest duration may be unlikely to provoke ASV false alarms, but can still fundamentally change the meaning of a given phrase\footnote{An example might be the manipulation of a phrase within a video posted to social media from \emph{`I won the election'} to \emph{`I lost the election'}.}.
Nonetheless, utterances in which a greater proportion of the utterance is bona fide still stand to fool the ASV system used in LA and PA tasks while also being more difficult to detect with a CM.
Depending of the application, given a partially spoofed utterance, it might then be of interest to estimate bona fide and spoofed labels at the segment level, 
i.e.\ to \emph{explain} the classifier decision. 

\textbf{Adversarial attacks} -- 
Many examples in the literature show the threat to reliable ASV stemming from adversarial attacks~\cite{villalba20_interspeech,Joshi2021}.
These target the ASV system, but rarely the CM. 
With estimates of security being only as strong as the adversary model, future editions of ASVspoof should consider adversarial attacks that target \emph{both} the ASV \emph{and} the CM. 

\textbf{DF metrics} -- Assessment in the case of LA and PA tasks reflects the tandem evaluation of separate ASV and CM subsystems.  Because there is no ASV system in the case of the DF task, assessment is applied to an independent CM system.  In practice, however, there \emph{is} a second sub-system, namely a human listener.  Since effective attacks must then fool both a CM and a human listener, there is no reason why some form of tandem assessment cannot also be applied for the DF task.  This might involve the combination of CM scores with some form of perceptual score, e.g.\ quality or speaker similarity.  Some form of objective scores, e.g.\ derived using MOSNet~\cite{mosnet}, might also be used as a proxy for subjective scores.

\section{Conclusions}
\label{Section:Conclusion}
The ASVspoof 2021 challenge was designed to foster progress in reliable automatic speaker verification, spoofing and deepfake detection in more realistic and practical scenarios. The challenge attracted submissions from 54 teams across the three tasks, each of whom submitted valid challenge scores, and 34 teams submitted the required system description. This paper provides an overview of the challenge datasets and an analysis of results. We also report an analysis of various data-related issues, in addition to new results for hidden data subsets.
They reveal some challenge limitations, prompting ideas for future challenge editions.

Results show that the transmission of speech data across real telephony systems in a logical access task causes only modest degradation to spoofing detection reliability and that estimates of performance from transmission across a local area network are as reliable as estimates from transmission across more geographically distant endpoints.
Impacts on performance caused by compression effects in a deepfake task are also modest, but results show a lack of generalisation to different source data.
Data augmentation is common to all top-performing systems for both logical access and deepfake tasks.  

The physical access task appears to be the most challenging of the three, likely due to the mismatch between training and evaluation data.
Results show the added difficulty when attacks are recorded with higher quality microphones and replayed with high-quality loudspeakers, both at a short distance from the talker and microphone respectively, or when the automatic speaker verification microphone is of lower quality.

Future editions of ASVspoof will continue the path towards ever-more realistic conditions and will consider stronger adversaries, larger databases containing greater variation in acoustic conditions and spoofing attacks, data collected from a larger speaker population and an additional, relaxed training data policy.  
Finally, ASVspoof and SASV challenges will merge to allow greater flexibility in the form of jointly-optimised CM and ASV sub-systems and alternative combination architectures.

\section*{Acknowledgments}
The ASVspoof 2021 organising committee extends its sincere gratitude and appreciation to all the 72 registered participants and teams, including those opted to not submit the challenge scores to the evaluation phase. Since participants are assured of anonymity, we regret that we cannot acknowledge them here by name. This work was supported jointly by the French National Research Agency (grant No. ANR-18-JSTS-0001) and the Japan Science and Technology Agency (grant No. JPMJCR18A6) under project VoicePersonae. The work also received partial support from Academy of Finland (proj. 349605, SPEECHFAKES) and MEXT KAKENHI Grants (21K17775, 21H04906). The work also received partial support from Region Grand Est, France, joint doctoral funding from Inria and UEF, and from the Science and Engineering Research Council, Agency for Science, Technology and Research (A$^\star$STAR), Singapore, through its Council Research Fund (Project No. CR-2021-005).

\section*{References}
{
\printbibliography
}

\newpage 

\section*{Supplementary material}
\begin{table}[h]
    \centering
    \caption{Details on microphone and replay devices used in PA data recording. Numbers are from official specifications.}
    \begin{tabular}{lccc}
    \toprule
    Replay device       & s2 & s3 & s4 \\
    \midrule
    \multirow{2}{*}{Brand} & SONY & NEUMANN & GENELC \\
          & SRS-XB43 & KH 80 DSP & 8030 CP \\
    \shortstack[l]{Lower cut-off \\ freq. (–6 dB) in Hz} & 20 & 53 & 47 \\
    \shortstack[l]{Upper cut-off \\ freq. (–6 dB) in Hz} & 20k & 22k & 25k \\
    \shortstack[l]{Passband free-field \\ FR ($\pm$ 2dB) in Hz} & UNAVAL. & 59 - 20k & 54 - 20k \\
    \bottomrule
    & \\
    \toprule
    Microphone       & M1/m1 & M2/m2 & M3/m3 \\
    \midrule
    \multirow{2}{*}{Brand}  & Marantz & M-Audio &  iPad Air\\
                            & MPM-1000 & Uber-mic & MEMS mic. \\
    FR in Hz &  20 - 20k & 30 - 20k & UNAVAIL. \\
    SNR  in dB & 77 & 110 & UNAVAIL. \\
    Sensitivity in dB & 45 & 38 ($\pm$ 2)& UNAVAIL.\\
    \bottomrule
    \end{tabular}
    \label{tab:PA_device}
\end{table}

\section{Devices used in PA data recording}
The replay and recording devices used to create PA data is shown in Table.~\ref{tab:PA_device}. All setups are cited from the specifications available at the official websites.

\section{Description of Top-Performing Systems}
The top-5 systems for each scenario are ranked and selected respectively based on minimum t-DCF. The architectures are presented in terms of block diagrams which were drawn based on the system descriptions and feedback from the challenge participants.

For a unified presentation, we break down each system into four cascaded processing blocks: 
\begin{itemize}
    \item \textbf{Audio pre-processor} includes the steps to partition data and process input wave files. Most of the methods in this step are applied only at training phase.
    \item \textbf{Input processor} contains the acoustic frontends used for further processing. Learnable acoustic frontends are also categorized to this part.
    \item \textbf{Classifier} covers the neural network as well as statistical models for the countermeasure system. Online augmentation methods and training strategies (if any) are also covered here.
    \item \textbf{Post-processor} contains score-domain operations, including score fusion and normalization.
\end{itemize}

\noindent In the following (starting from next page) we give short descriptions to the architecture of the illustrated systems.

\clearpage
\newpage

\begin{figure*}[t]
    \centering
    \includegraphics[width=1.0\textwidth]{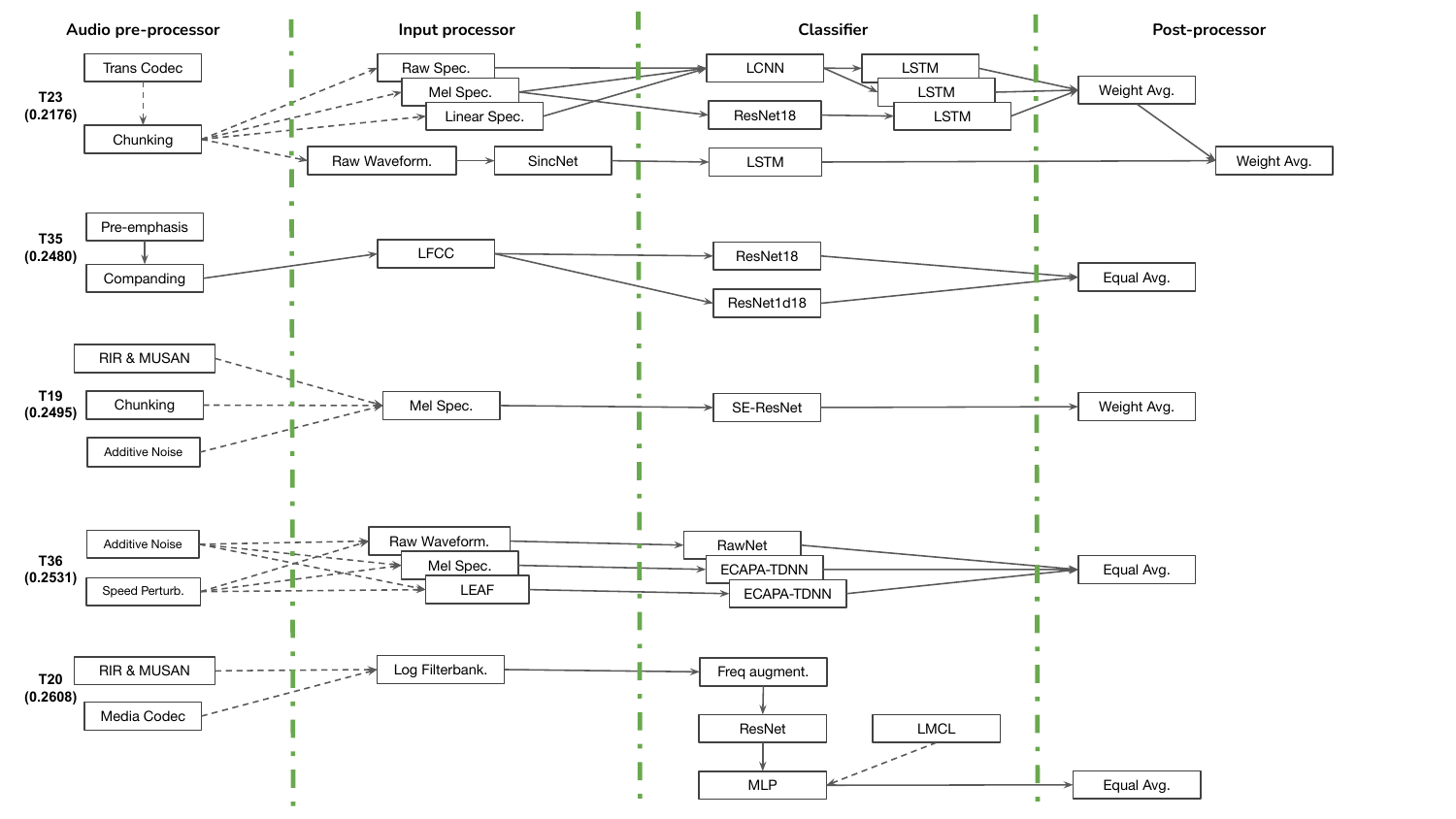}
        \caption{Illustration of top-5 systems for the LA scenario. The descending order is based on performance via minimum t-DCF values on evaluation set. Numbers in brackets are the values. Dash lines indicate steps/modules which are only applied for training, while solid lines indicate ones applied for both training and testing.}
    \label{fig:LA}
\end{figure*}

\subsection{Logical Access (LA)}
\textbf{T23}: Multiple sub-systems are trained based on the codec-augmented and trimmed audio. Most of them are based on various types of spectral features and two parallel classifiers, LightCNN (LCNN) and ResNet. They are forwarded to LSTMs and the scores are summed up with weights. Further, another sub-system uses a SincNet frontend and an LSTM backend. The output score of this subsystem is further fused with the weighted averaged score above.

\textbf{T35}: The input audio is processed using pre-emphasis and a-law companding algorithm. LFCC features are extracted from the processed audio and fed into two ResNet-based classifiers. The output scores from the two classifiers are averaged.

\textbf{T19}: The training data is augmented via RIR and MUSAN datasets, followed by trimming and additive noise. The mel spectrograms are fed into a ResNet with squeeze-and-excitation in its building blocks. The output scores are averaged via pre-defined empirical weights.

\textbf{T36} is a combination of multiple classifiers based on RawNet and ECAPA-TDNN. The training data is augmented via additive noise and 3-fold speed perturbation with parameters of 0.9, 1.0 and 1.1. Three types of feature extractors are used --- raw waveform, mel spectrogram and a learnable LEAF frontend, whose outputs are fed into one RawNet and two ECAPA-TDNNs. The output scores are combined using equal weights.

\textbf{T20} uses a ResNet classifier trained with log filterbanks. The training data is augmented via RIR, MUSAN and media codecs such as mp3 and m4a. The log filterbanks are augmented via zero-value masking at frequency domain (FreqAugment). A fully-connected layer is added after the ResNet and trained with \emph{large margin cosine loss} (LMCL).

\clearpage
\newpage

\begin{figure*}[t]
    \centering
    \includegraphics[width=1.0\textwidth]{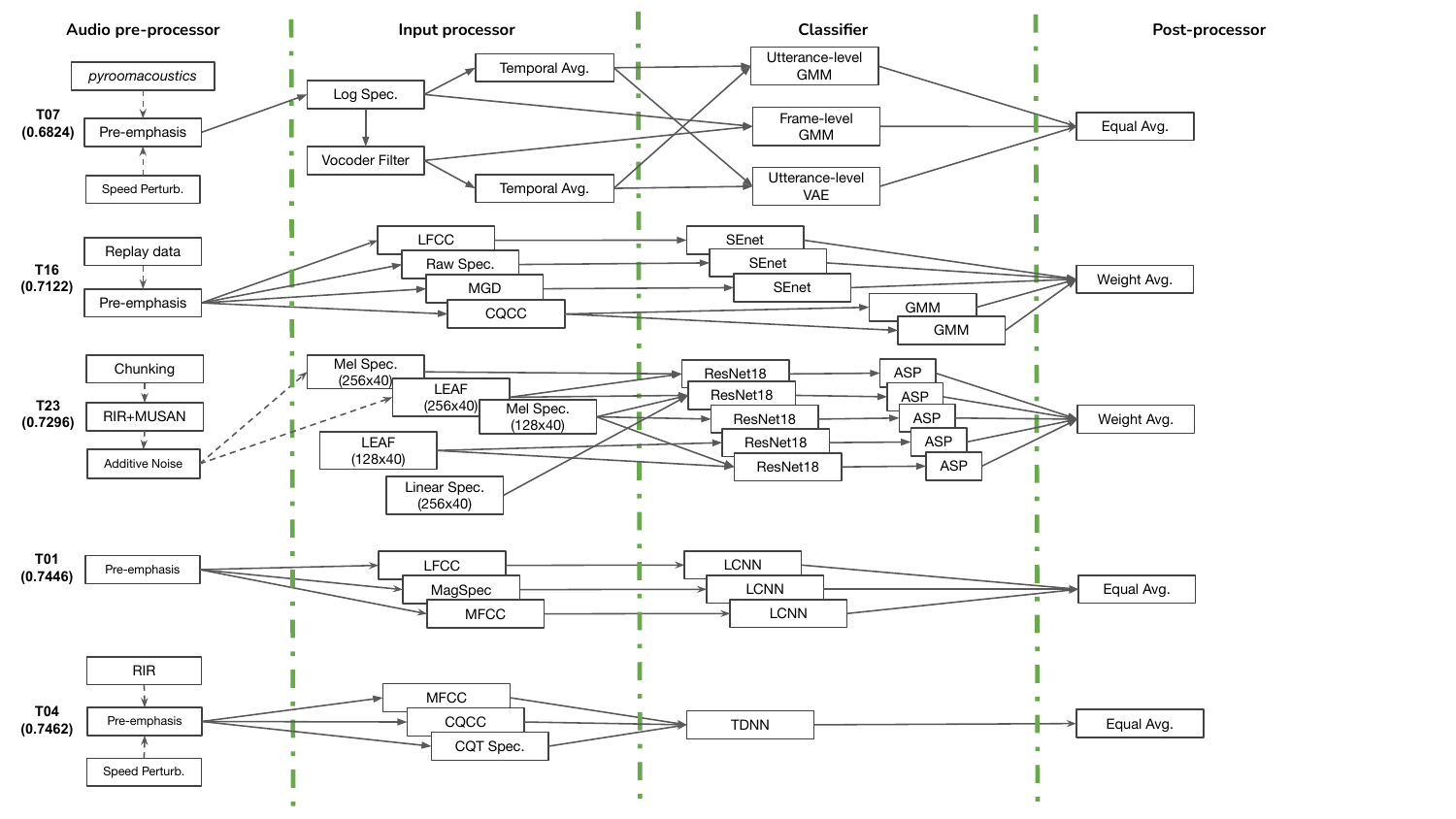}
    \caption{Illustration of top-5 systems for the PA scenario. The descending order is based on performance via minimum t-DCF values on evaluation set. Numbers in brackets are the values.  Dash lines indicate steps/modules which are only applied for training, while solid lines indicate ones applied for both training and testing.}
    \label{fig:PA}
\end{figure*}

\subsection{Physical Access (PA)}
\textbf{T07} is an one-class learning framework based on \emph{Gaussian mixture models} (GMM) and \emph{variation autoencoder} (VAE). The training data is augmented via room simulation \footnote{via \emph{pyroomacoustics}: \url{https://github.com/LCAV/pyroomacoustics}} and 2-fold speed perturbation with parameters of 0.9 and 1.1. The acoustic feature used is log spectrogram, which is fed into a neural vocoder to create filtered spectrograms. Applied candidates for the vocoder are WORLD, HifiGAN, and MelGAN. The original and processed spectrograms, which are time-frequency representations, are averaged via the temporal axis. One GMM and one VAE are trained via the temporal-averaged spectrograms and another GMM is trained via the original log spectrogram and the filtered spectrogram. The output scores are combined via equally weighting.

\textbf{T16} uses a combination of SEnet and GMM trained with various short-term spectral features. The data is augmented via in-house replay data. The scores are fused using empirically weighted averaging.

\textbf{T23}: A parallel combination of multiple sub-systems trained on trimmed speech segments, augmented via RIR, MUSAN and other additive noises. Mel spectrogram with various numbers of frequency bins and LEAF frontend are the feature extractors. Classifiers are implemented via ResNet, whose output are forwarded to scors via \emph{attentive statistical pooling} (ASP). Final score is obtained via empirically weight averaging.

\textbf{T01}: The pre-emphasized audio is input to produce three different types of acoustic features: LFCCs, magnitude spectrogram (MagSpec), and MFCCs. Each of them is fed into a LightCNN classifier and the output score is obtained by fusing the output of the three classifiers using equal weights.

\textbf{T04}: A TDNN network with statistical pooling. The pre-emphasized audio is used to produce another three types of features: MFCCs, CQCCs and CQT spectrogram.

\clearpage
\newpage
\begin{figure*}[t]
    \centering
    \includegraphics[width=1.0\textwidth]{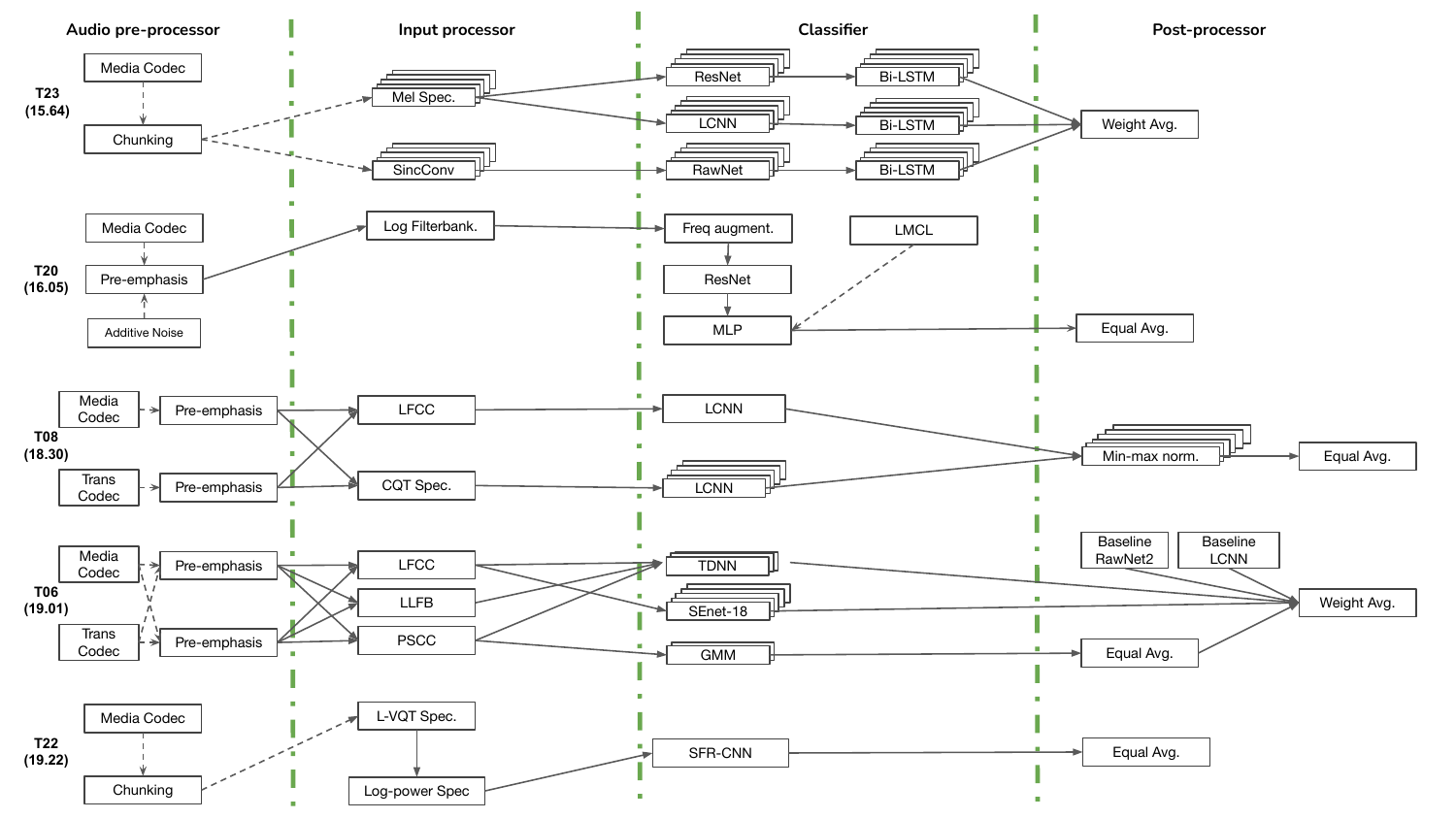}
    \caption{Illustration of top-5 systems for the DF scenario. The descending order is based on performance via equal error rates on evaluation set. Numbers in brackets are the error rates.  Dash lines indicate steps/modules which are only applied for training, while solid lines indicate ones applied for both training and testing.}
    \label{fig:DF}
\end{figure*}

\subsection{DeepFake (DF)}
\textbf{T23} is a parallel hybrid of three classifiers. Two acoustic frontends process the trimmed and codec-augmented audio: Mel spectrogram extractor and a convolutive frontend built via sinc function. The former is fed into a ResNet and a LightCNN and the latter is fed into a RawNet. The outputs of the three nets is feed into separate bi-directional LSTMs to produce scores, which are finally fused via empirically weight averaging.

\textbf{T20}: The feature extractor, classifier and scoring scheme is the same as one used by the same team in the LA scenario. The training data is augmented via media codecs such as mp3 and m4a, and other additive noises.

\textbf{T08} fuses multiple LightCNN classifiers using LFCCs and CQT spectrograms. The input data is augmented using multiple media and transmission codecs\footnote{G722, alaw, G723, G726, gsm, opus, SPEEX, mp3, m4a, ogg, mp2}. LFCCs are used to train one LightCNN and the remaining classifiers are trained on the other feature. The scores are processed via min-max normalization, before being averaged. 

\textbf{T06} is a combination of a TDNN classifier trained on LFCCs, four SEnet modules trained on LFCCs, two GMMs trained on \emph{product spectral cepstral coefficients} (PSCCs), and three TDNNs trained separately on the two mentioned features and \emph{log linear filterbank energies} (LLFBs).

\textbf{T22} is a convolutional network based on CondenseV2Net, trained on log power spectrum derived from \emph{long-term variable Q transform} (L-VQT). The training data is augmented via mp3, m4a, wma and trimmed to 4-second chunks.

\clearpage
\newpage 

\begin{figure*}[t]
\centering
\subfloat[LA]{
\includegraphics[height=4cm]{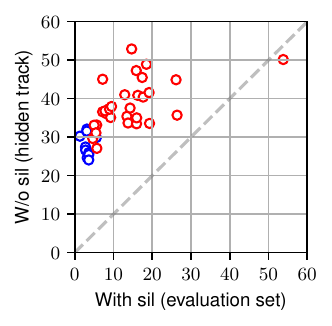}
}
\subfloat[PA (hidden track 1)]{
\includegraphics[height=4cm]{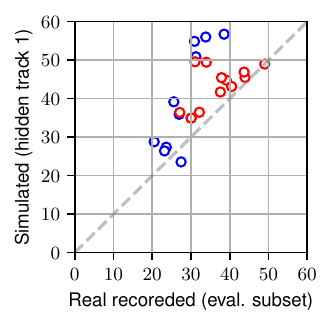}
}
\subfloat[PA (hidden track 2)]{
\includegraphics[height=4cm]{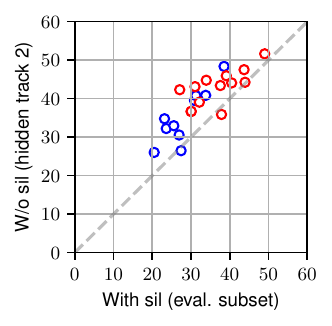}
}
\subfloat[DF]{
\includegraphics[height=4cm]{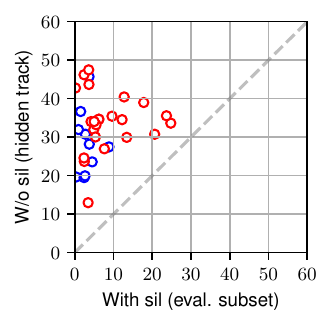}
}
\caption{Comparison of CMs EERs on evaluation set and hidden tracks. Blue dots correspond to top 1-10 submissions. \textcolor{red}{Red dots} are other submissions. Best viewed in color.}
\label{fig:hidden_track}%
\end{figure*}

\begin{table*}[t!]
    \centering
    \vspace{-5mm}
    \setlength{\tabcolsep}{3pt}
    \caption{Experiment on non-speech regions using official baselines.}
    \resizebox{\textwidth}{!}{
    \begin{tabular}{|c|l|c|c|c|c|c|c|c|c|}
    \hline
         \multicolumn{2}{|c|}{}  & \multicolumn{2}{c|}{B01} & \multicolumn{2}{c|}{B02} & \multicolumn{2}{c|}{B03} & \multicolumn{2}{c|}{B04} \\ \cline{3-10}
         \multicolumn{2}{|c|}{}  & EER (\%) & min t-DCF & EER (\%) & min t-DCF  & EER (\%) & min t-DCF  & EER (\%) & min t-DCF \\ \hline
       \multirow{4}{*}{\shortstack{LA}}   
                             & w/ non-speech (eval. set) & 15.62 &  0.497 &  19.30 &  0.576 &  9.26 &  0.345 &  9.50 &  0.426\\  \cline{2-10}
                             & w/o non-speech between words &   14.84 &  0.567 &  19.17 &  0.642 &  8.82 &  0.460 &  7.80 &  0.471 \\ \cline{2-10}
                             & w/o non-speech at two ends & 34.48 &  0.858 &  33.68 &  0.849 &  34.02 &  0.926 &  35.00 &  0.908 \\ \cline{2-10}
                             & w/o any non-speech (hidden track) & 34.39 &  0.862 &  33.57 &  0.842 &  35.10 &  0.929 &  37.97 &  0.961 \\ \hline\hline
       \multirow{4}{*}{\shortstack{PA}}   
                             & with non-speech  (eval. subset) & 30.02 &  0.823 &  32.16 &  0.883 &  49.02 &  1.000 &  43.95 &  0.962\\  \cline{2-10}
                             & w/o non-speech between words & 29.77 &  0.839 &  32.71 &  0.906 &  48.84 &  1.000 &  43.23 &  0.959 \\ \cline{2-10}
                             & w/o non-speech at two ends & 31.22 &  0.862 &  37.13 &  0.991 &  50.09 &  1.000 &  40.08 &  0.966 \\ \cline{2-10}
                             & w/o any non-speech (hidden track) & 36.65 &  0.959 &  39.09 &  0.997 &  51.65 &  1.000 &  44.26 &  0.998 \\ \hline\hline
       \multirow{4}{*}{\shortstack{DF}}   
                             &  with non-speech (eval. subset) & 20.62 &   -    &  24.79 &  -     &  12.22 & -      &  6.22 &  - \\  \cline{2-10}
                             & w/o non-speech between words &19.18 & - & 23.58 & -& 11.16 &   -    &  4.85 &  -    \\  \cline{2-10}
                             & w/o non-speech at two ends &30.71 & -& 33.33 & -& 34.38 & -      &  32.32 & -      \\  \cline{2-10}
                             & w/o any non-speech (hidden track)  & 30.76 &     -  &  33.64 &   - & 34.54 &  -     &  34.65 &   -   \\  \hline
    \end{tabular}
    }
    \label{tab:exp_nonspeech}
\end{table*}

\begin{figure*}[h!]
  \centering
  \subfloat{
    \centering
    \includegraphics[width=\textwidth]{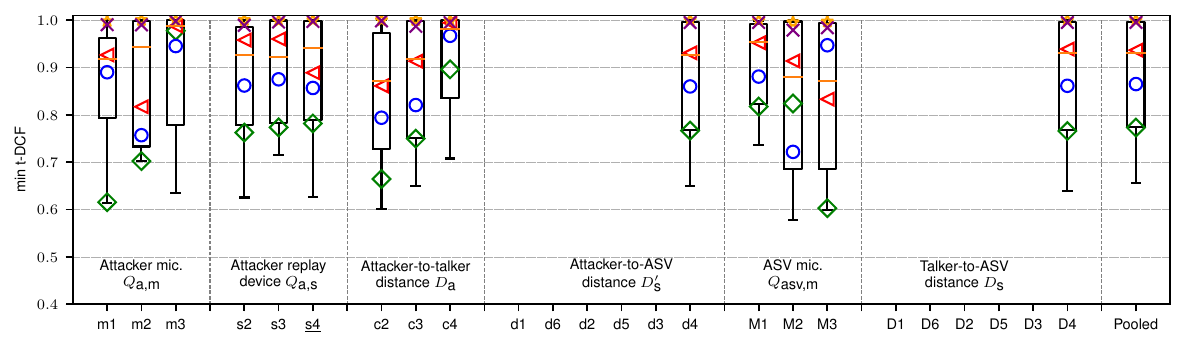}
    \label{fig:pa_decomposed_eer_hidden_subset}
  }
  \vspace{-5mm}
  \caption{Boxplots of min t-DCF of top-10 submissions decomposed over different factors on PA simulated subset (hidden track 1). Markers are top-1 submission} (\textcolor{teal}{$\diamond$}), B01 (\textcolor{blue}{o}), B02 (\textcolor{red}{$\triangleleft$}), B03 (\textcolor{orange}{$\medwhitestar$}), and B04 (\textcolor{purple}{$\times$}). Note that data in PA hidden track 1 only covers $D_s' = $d4 and $D_s=$D4.
  \label{fig:box_PA_eval_hidden}
\end{figure*}

\begin{table*}[h!]
    \centering
    \setlength{\tabcolsep}{6pt}
    \caption{Min t-DCFs of top-5 CMs in evaluation subset ($\largewhitestar$) and simulated hidden track ($\largeblackstar$)}. 
    \resizebox{\textwidth}{!}{
    \begin{tabular}{lcccccccccccccccc}
    \toprule
        & & \multicolumn{3}{c}{${Q}_{\mbox{a,m}}$} & \multicolumn{3}{c}{${D}_{\mbox{a}}$} &  \multicolumn{3}{c}{${Q}_{\mbox{a,s}}$} &   ${D}_{\mbox{s}}'$ &  \multicolumn{3}{c}{${Q}_{\mbox{asv,m}}$} & ${D}_{\mbox{s}}$ &\\
    \cmidrule(lr){3-5}\cmidrule(lr){6-8}\cmidrule(lr){9-11}\cmidrule(lr){13-15}
        &                         &           m1            &           m2            &           m3            &           s2            &           s3            & $\underline{\mbox{s4}}$ &           c2            &           c3            &           c4            &              d4            &           M1            &           M2            &           M3            &                 D4            &         Pooled         \\ 
    \midrule
        T07 $\largewhitestar\largeblackstar$   &real recorded&         0.684         &\textbf{0.655}&\textbf{0.498}&\textbf{0.535}&\textbf{0.667}&\textbf{0.636}&\textbf{0.447}&\textbf{0.606}&\textbf{0.765}&\textbf{0.607}&\textbf{0.450}&\textbf{0.402}&         0.656         &\textbf{0.597}&\textbf{0.614}\\
   &simulated&\textbf{0.615}&         0.702         &         0.978         &         0.763         &         0.774         &         0.782         &         0.664         &         0.750         &         0.896         &         0.767         &         0.818         &         0.824         &\textbf{0.603}&         0.766         &         0.773         \\ \midrule
T16 $\largewhitestar$   &real recorded&\textbf{0.871}&\textbf{0.782}&\textbf{0.771}&\textbf{0.730}&\textbf{0.848}&\textbf{0.840}&\textbf{0.720}&\textbf{0.786}&\textbf{0.907}&\textbf{0.803}&\textbf{0.823}&\textbf{0.725}&\textbf{0.707}&\textbf{0.803}&\textbf{0.809}\\
   &simulated&         0.962         &         0.999         &         1.000         &         0.977         &         1.000         &         0.999         &         0.973         &         0.999         &         1.000         &         0.998         &         0.996         &         0.998         &         0.994         &         0.998         &         0.998         \\ \midrule
T23 $\largewhitestar\largeblackstar$   &real recorded&\textbf{0.718}&\textbf{0.750}&\textbf{0.569}&\textbf{0.562}&\textbf{0.673}&\textbf{0.786}&\textbf{0.645}&\textbf{0.675}&\textbf{0.716}&\textbf{0.672}&\textbf{0.637}&\textbf{0.434}&         0.686         &\textbf{0.668}&\textbf{0.680}\\
   &simulated&         0.780         &         0.776         &         0.778         &         0.720         &         0.804         &         0.810         &         0.803         &         0.759         &         0.773         &         0.771         &         0.939         &         0.634         &\textbf{0.647}&         0.774         &         0.779         \\ \midrule
T01 $\largewhitestar$  &real recorded&\textbf{0.850}&\textbf{0.778}&\textbf{0.802}&\textbf{0.729}&\textbf{0.795}&\textbf{0.895}&\textbf{0.794}&\textbf{0.804}&\textbf{0.831}&\textbf{0.805}&\textbf{0.841}&\textbf{0.683}&\textbf{0.807}&\textbf{0.806}&\textbf{0.811}\\
   &simulated&         0.963         &         0.998         &         1.000         &         0.989         &         0.999         &         0.992         &         0.970         &         0.998         &         1.000         &         0.996         &         0.986         &         0.998         &         0.994         &         0.996         &         0.996         \\ \midrule
T04 $\largewhitestar\largeblackstar$   &real recorded&         0.758         &         0.806         &\textbf{0.600}&\textbf{0.604}&\textbf{0.711}&         0.832         &         0.676         &         0.734         &         0.751         &         0.717         &\textbf{0.709}&\textbf{0.542}&         0.665         &         0.709         &         0.722         \\
   &simulated&\textbf{0.613}&\textbf{0.718}&         0.634         &         0.625         &         0.715         &\textbf{0.626}&\textbf{0.601}&\textbf{0.649}&\textbf{0.708}&\textbf{0.650}&         0.735         &         0.578         &\textbf{0.600}&\textbf{0.640}&\textbf{0.656}\\ \midrule
T27 $\largeblackstar$  &real recorded&\textbf{0.721}&         0.728         &\textbf{0.576}&\textbf{0.650}&\textbf{0.699}&         0.679         &\textbf{0.645}&\textbf{0.658}&\textbf{0.717}&\textbf{0.668}&\textbf{0.690}&\textbf{0.538}&\textbf{0.727}&\textbf{0.664}&\textbf{0.676}\\
   &simulated&         0.833         &\textbf{0.719}&         0.688         &         0.825         &         0.775         &\textbf{0.653}&         0.704         &         0.713         &         0.817         &         0.744         &         0.761         &         0.641         &         0.800         &         0.743         &         0.752         \\ \midrule
T03 $\largeblackstar$   &real recorded&\textbf{0.888}&         0.848         &\textbf{0.759}&\textbf{0.777}&\textbf{0.843}&         0.876         &         0.824         &         0.843         &\textbf{0.836}&\textbf{0.827}&\textbf{0.862}&\textbf{0.634}&\textbf{0.823}&\textbf{0.833}&\textbf{0.835}\\
   &simulated&         0.893         &\textbf{0.760}&         0.947         &         0.864         &         0.877         &\textbf{0.859}&\textbf{0.797}&\textbf{0.823}&         0.969         &         0.862         &         0.882         &         0.723         &         0.951         &         0.864         &         0.867         \\ 
    \bottomrule
    \end{tabular}
    }
    \label{tab:pa_decomposed_eer_hidden_subset}
    Note: for a fair comparison, the real recorded evaluation subset only covers data in $D'_{s}=\text{d4}$ and $D_s=\text{D4}$. 
\end{table*}

\section{Hidden Track}

\subsection{Hidden track results from challenge participants}
We recall that the main goal of hidden track data is to assess the dependence of CM upon certain data characteristics such as non-speech and real/simulated attacks for PA track. The comparison of EERs for evaluation set and hidden track are presented in Fig.~\ref{fig:hidden_track}. We can see that for LA and DF scenarios, while for the main track the EERs for the top systems are all lower than 10\%, for the hidden track the number raised to more than 20\%. Such pattern can also be found for the PA scenario, with narrower performance gap between the main and hidden tracks. 

The results indicate the effect of silence on the spoofing detection. Meanwhile, the different gap between PA and the other two scenarios may due to the artifact on the length of silence injected in the spoofed and bonafide training data --- for LA and DF, where synthetic speeches are produced via TTS systems, such difference is large. For PA where the attack comes from replaying the audio, such difference becomes smaller. 

\subsection{Additional analysis on role of non-speech}
The hidden track data has all the non-speech intervals removed, including those at the two ends of the utterances (i.e., leading and trailing non-speech) and others between speech sounds. To further investigate which type of non-speech affects the CM performance, we prepared two additional test sets that cover the same set of utterances as the official hidden track. One set only removes the non-speech at the two ends, while the other only removes the non-speech intervals between words. Post-processing based on an energy-driven VAD was employed to accommodate low-energy segments. Specifically, we classify frames of non-speech as speech if non-speech intervals occurring between words is less than 50 milliseconds in duration. As a result, this ensures that nasals, or voiceless plosive consonants are incorporated as segments of speech.

The four official baselines are used to score the two data sets, and the results are listed in Table~\ref{tab:exp_nonspeech}. For reference, results on the official hidden track and evaluation data\footnote{On the PA task, the official hidden track only contains data with a talker-to-ASV distance equal ($D_s$) to D4 and an attacker-to-ASV distance ($D_s'$) equal to d4. Hence, the data with non-speech is a subset of the evaluation set with $D_s=$D4 or $D_s'=$d4. Similarly, the data with non-speech on DF task has the same data source as the hidden subset. In contrast, the evaluation data with non-speech for LA is equal to the official evaluation set.} are added to the table.
Results on the LA and DF tasks show that the non-speech intervals at the beginning and the end of the utterances provide more information than those between words. Trimming the non-speech intervals at the two ends made the utterances much more difficult to detect. For example, B04 on the LA task obtained a min t-DCF value of 0.426 on the evaluation data with non-speech. Trimming the non-speech between words led to a slightly higher min t-DCF 0.471, but trimming those at the two ends led to a much higher value of 0.908, which is similar to the min t-DCF of 0.961 when all non-speech regions were trimmed.

On the PA task, the results of B01 and B02 also showed that the non-speech regions at the two ends degraded the EER and min t-DCF, but the differences across the four data sets are smaller compared to those on the LA and DF tasks.

\subsection{Additional analysis on PA simulated hidden track}

We found that most of the CMs performed worse on the simulated hidden track than on the evaluation data (Sec. IV.B of the paper). We computed the min t-DCFs under different simulated rooms and devices and plotted the results in Fig.~\ref{fig:box_PA_eval_hidden}. Compared with those on the evaluation data (Fig. 4b in the paper), the median min t-DCF is more similar across different simulated rooms, devices, and other factors.

The simulated data was created from impulse responses, which were estimated from sweep signals recorded by various devices in rooms used for the real data recording. The impulse responses are expected to capture the room acoustics and frequency response characteristics of the devices. However, the estimation is not error-free, and the estimated impulse responses may not accurately reflect different room acoustics and frequency response characteristics of different devices. This may be one reason for the higher min t-DCF medians across different factors. Useful artefacts that discriminate real recorded bona fide from real replayed spoofed data may have been lost during the simulation process.

However, not all the CMs performed worse on the simulated data.  Table~\ref{tab:pa_decomposed_eer_hidden_subset} shows the min t-DCF of the top-5 CMs on the evaluation set (which are marked with $\largewhitestar$) and the top-5 CMs on the hidden track ($\largeblackstar$). It is interesting to note that the min tDCF of T16 and T01 increased to around 1.0 for most of the conditions on the simulated data, even though they are among the top-5 on the real recorded evaluation data. In contrast, T07 and T16 saw less degradation, and T04 even performed better. One notable difference is that T07, T16, and T04 used room-impulse-based data augmentation, while T16 and T10 did not (see Fig.\ref{fig:PA}). This suggests that with certain training strategies, the CMs can do well on both the real and simulated data.

\clearpage
\newpage

\begin{table*}[th]
    \centering
    \caption{Results of the investigation on different frontends, backend classifiers and training losses. The evaluation metrics are EER and minimum t-DCF.}
\begin{tabular}{|r|c|c|c|c|c|c|c|c|}
    \hline
          \multirow{2}{*}{ID} & \multirow{2}{*}{Front end} & \multirow{2}{*}{Back end} & Training & \multicolumn{2}{|c|}{\emph{2019 LA}} & \multicolumn{2}{|c|}{\emph{2021 LA}} & \emph{2021 DF}  \\ \cline{5-9}
          &   &  &  criterion & EER(\%) & t-DCF & EER(\%) & t-DCF & EER(\%) \\ \hline
         \texttt{M01} & LFCC & \multirow{3}{*}{\shortstack{LCNN-LSTM GAP}} & \multirow{5}{*}{\shortstack{AM-softmax}}  & 3.43 & 0.079 & 22.68 & 0.765 & 27.13 \\ \cline{1-2}\cline{5-9}
         \texttt{M02} & Spec. &  &  & 5.96	& 0.134 & \textbf{12.08} & \textbf{0.418} & \textbf{18.05} \\ \cline{1-2}\cline{5-9}
         \texttt{M03} & LFB & &  & 6.23 & 0.199 & 29.95 & 0.871 & 32.14 \\ \cline{1-3}\cline{5-9}
         \texttt{M04} & \multirow{5}{*}{\shortstack{LFCC}} & LCNN-LSTM SAP & & 3.93 & 0.075 & 23.13 & 0.698 & 25.21 \\ \cline{1-1}\cline{3-3}\cline{5-9}
         \texttt{M05} & & LCNN-LSTM FF & & 3.04 & 0.068 & 22.73 & 0.758 & 28.29 \\ \cline{1-1}\cline{3-5}\cline{5-9}
         \texttt{M06} & & \multirow{3}{*}{\shortstack{LCNN-LSTM GAP}}  & OC-softmax & 2.96 & 0.076 & 21.72 & 0.613 & 24.95 \\ \cline{1-1}\cline{4-9}
         \texttt{M07} & & & P2SGrad & \textbf{2.62} & \textbf{0.067} & 19.77 & 0.628 & 34.40 \\ \cline{1-1}\cline{4-9}
         \texttt{M08} & & & Vanilla Softmax & 3.19 & \textbf{0.067} & 22.52 & 0.766 & 26.74 \\ \hline
    \end{tabular}
    \label{tab:results_frontends}
\end{table*}

\begin{table*}[th]
    \centering
    \caption{Results of applying data augmentation onto the training data. The model evaluated for this set of experiments is \texttt{M02} in Table \ref{tab:results_frontends}.}
        \begin{tabular}{|c|c|c|c|c|c|}
    \hline
          & \multicolumn{2}{|c|}{\emph{2019 LA}} & \multicolumn{2}{|c|}{\emph{2021 LA}} & \emph{2021 DF}  \\ \hline
         Augmentation & EER(\%) & t-DCF & EER(\%) & t-DCF & EER(\%) \\ \hline  
         (No augmentation) & 5.96 & 0.134 & 12.08 & 0.418 & 18.05 \\ \hline
         \texttt{NOISE} & 5.30 & 0.156 & 11.81 & 0.402 & 19.79 \\ \hline
         \texttt{MP3} & 4.54 & 0.129 & 11.48 & 0.394 & 19.69 \\ \hline
         \texttt{REVERB} & 5.30 & 0.141 & 12.43 & 0.415 & 18.87 \\ \hline
    \end{tabular}
    \label{tab:results_da}
\end{table*}

\section{Analysis on Practicality of Common Techniques}
We here provide a practical, controlled analysis using baseline-level systems. We investigate the effect of three factors on the spoofing detection performance --- feature extractors, backend classifiers, and loss functions. The feature extractors acquired here are linear filterbanks (LFB), linear frequency cepstral coefficients (LFCCs), and spectrogram. Those features are widely acquired for the submitted systems, according to the system descriptions and feedback metadata. The backend classifier is based on light convolutional neural networks (LCNN), but with fixed-size input trim-and-pad strategy, global average pooling (GAP), or self-attentive pooling (SAP) layer. The training loss functions varies among cross-entropy over additive margin (AM) softmax, one-class (OC) softmax, vanilla softmax, and mean square error with Probability-to-Similarity Gradient (P2SGrad). We follow the training scheme described in the manuscript for LA and DF scenarios and evaluate the trained systems on 2019 LA, 2021 LA and 2021 DF datasets. 

Results are presented in Table \ref{tab:results_frontends} with \texttt{M01} as the baseline. We further conduct investigation on the effect of three data augmentation methods is conducted based on the detailed information of data augmentation techniques from the participants, using the best-performed on 2021 LA and DF reported in Table \ref{tab:results_frontends} (\texttt{M02}). We selected some most commonly-used ones and analyze their impact on the performance, namely \emph{room impulse response}, MUSAN noise, and mp3 compression. We make the following observations:
    \begin{itemize}
        \item Comparing \texttt{M01} --- \texttt{M03}, switching the frontend features from LFCC to spectrogram degrades the performance on 2019 LA. But \texttt{M02} returns the best performance on the two 2021 datasets among all systems.
        \item Concerning \texttt{M06} --- \texttt{M08}, best performance on 2019 LA on both metrics is obtained by switching the training loss from AM-softmax to P2SGrad (\texttt{M07}), but the same system returned worse performance on 2021 DF.
        \item Regarding data augmentation methods whose performance are presented in Table \ref{tab:results_da}, all applied return better performance on 2019 LA as expected in terms of EER, and 2021 LA in terms of minimum t-DCF. However, on 2021 DF, best performance is obtained without any augmentation applied.
    \end{itemize}
Based on these observations, it appears that more detailed feature representations (e.g., raw spectrograms) have an edge over filterbank-integrated spectral representations (e.g., LFCCs and LFBs), especially in more challenging data conditions. The choice of the loss function is critical as well, even if the findings are inconsistent across datasets. Regarding data augmentation, the challenge participants applied many data augmentation techniques. Despite our initially hopeful expectations, we only observe modest improvements in our post-hoc analysis. The likely reason for the differences lies in the implementation details of the data augmentations used by the participants. For future editions of ASVspoof, it might be recommended for the participants to share the details of their data augmentation pipeline for reproducibility.


\section{Challenge Overall Results}

We present full numerical results on progress and evaluation sets in Table \ref{tab:LA_results}, \ref{tab:PA_results} and \ref{tab:DF_results}, for LA, PA and DF scenarios respectively. Same as last section, the team IDs are kept consistent with ones used in the main paper. The IDs starting with `B' are IDs of the baseline systems provided by us, for comparability concern. For LA and PA we present both t-DCF and EER results, while for DF we report EERs. Detailed description and analysis on the results are presented in our workshop paper\footnote{\url{https://www.isca-speech.org/archive/asvspoof_2021/yamagishi21_asvspoof.html}}.

\begin{table}[h]
    \centering
    \caption{ASVspoof 2021 progress and evaluation results for LA. Results shown in terms of pooled normalised minimum t-DCF and pooled EER [\%].}
    \begin{tabular}{cccccc}
    \toprule
    
           &          &    \multicolumn{2}{c}{Progress set} & \multicolumn{2}{c}{Evaluation set} \\
    \cmidrule(lr){3-4}\cmidrule(lr){5-6}
       \#  & ID       &     t-DCF   &  EER      & t-DCF   &  EER       \\ 
    \midrule
         1   &   T23 &   0.1815 &  0.89  &  0.2177 &  1.32 \\ 
         2   &   T35 &   0.2115 &  2.61  &  0.2480 &  2.77 \\ 
         3   &   T19 &   0.2174 &  2.69  &  0.2495 &  3.13 \\ 
         4   &       &   0.2119 &  2.51  &  0.2500 &  2.81 \\ 
         5   &   T36 &   0.2373 &  3.69  &  0.2531 &  3.10 \\ 
         6   &   T20 &   0.2137 &  2.39  &  0.2608 &  3.21 \\ 
         7   &   T08 &   0.2376 &  3.23  &  0.2672 &  3.62 \\ 
         8   &   T16 &   0.2393 &  3.39  &  0.2689 &  3.63 \\ 
         9   &       &   0.2435 &  3.11  &  0.2725 &  3.61 \\ 
        10   &   T04 &   0.2371 &  4.54  &  0.2747 &  5.58 \\ 
        11   &   T06 &   0.2475 &  5.61  &  0.2853 &  5.66 \\ 
        12   &       &   0.2556 &  4.22  &  0.2880 &  5.01 \\ 
        13   &   T03 &   0.2461 &  3.65  &  0.2882 &  4.66 \\ 
        14   &       &   0.2556 &  5.90  &  0.2893 &  5.70 \\ 
        15   &   T31 &   0.2621 &  5.00  &  0.3094 &  5.46 \\ 
        16   &   T17 &   0.2989 &  7.10  &  0.3279 &  7.19 \\ 
        17   &   T07 &   0.3029 &  7.88  &  0.3310 &  8.23 \\ 
        18   &   T30 &   0.2998 &  7.58  &  0.3362 &  8.89 \\ 
        19   &   B03 &   0.3152 &  8.90  &  0.3445 &  9.26 \\ 
        20   &   T02 &   0.3219 &  7.71  &  0.3446 &  7.79 \\ 
        21   &   T14 &   0.3167 &  8.82  &  0.3451 &  8.98 \\ 
        22   &   T11 &   0.3015 &  6.80  &  0.3666 &  7.19 \\ 
        23   &   T34 &   0.3688 &  13.25 &  0.4059 &  13.45\\ 
        24   &   B04 &   0.4152 &  9.49  &  0.4257 &  9.50 \\ 
        25   &   T15 &   0.2986 &  9.67  &  0.4890 &  14.68\\ 
        26   &   B01 &   0.4948 &  15.80 &  0.4974 &  15.62\\ 
        27   &   T25 &   0.2086 &  2.03  &  0.5148 &  13.75\\ 
        28   &   T32 &   0.4256 &  10.93 &  0.5270 &  12.90\\ 
        29   &       &   0.5887 &  19.04 &  0.5748 &  18.50\\ 
        30   &   B02 &   0.5836 &  21.13 &  0.5758 &  19.30\\ 
        31   &       &   0.2528 &  5.26  &  0.5775 &  14.28\\ 
        32   &   T01 &   0.2772 &  6.32  &  0.6204 &  15.95\\ 
        33   &       &   0.2792 &  7.87  &  0.6288 &  15.87\\ 
        34   &   T24 &   0.3108 &  10.85 &  0.6320 &  15.98\\ 
        35   &       &   0.3233 &  9.72  &  0.6371 &  16.27\\ 
        36   &   T29 &   0.3316 &  11.87 &  0.6741 &  17.41\\ 
        37   &       &   0.3678 &  12.54 &  0.6813 &  17.66\\ 
        38   &   T12 &   0.6900 &  26.26 &  0.7228 &  26.41\\ 
        39   &       &   0.4381 &  13.77 &  0.7233 &  19.19\\ 
        40   &       &   0.6123 &  23.16 &  0.8521 &  26.14\\ 
        41   &       &   0.9988 &  53.16 &  1.0000 &  53.81\\ 
    \bottomrule
    \end{tabular}
    \label{tab:LA_results}
\end{table}

\begin{table}[h]
    \centering
    \caption{ASVspoof 2021 progress and evaluation results for PA. Results shown in terms of pooled normalised minimum t-DCF and pooled EER [\%].}
    \begin{tabular}{cccccc}
    \toprule
    
           &          &    \multicolumn{2}{c}{Progress set} & \multicolumn{2}{c}{Evaluation set} \\
    \cmidrule(lr){3-4}\cmidrule(lr){5-6}
       \#  & ID       &     t-DCF   &  EER      & t-DCF   &  EER       \\ 
    \midrule
    1   & T07 &     0.6736 &  23.60 &  0.6824 &  24.25\\ 
    2   & T16 &     0.6790 &  25.37 &  0.7122 &  27.59\\ 
    3   & T23 &     0.7019 &  24.22 &  0.7296 &  26.42\\ 
    4   & T01 &     0.6925 &  25.73 &  0.7446 &  28.36\\ 
    5   & T04 &     0.6676 &  24.80 &  0.7462 &  29.00\\ 
    6   &     &     0.7207 &  27.23 &  0.7469 &  29.22\\ 
    7   & T33 &     0.7263 &  27.00 &  0.7648 &  29.55\\ 
    8   & T08 &     0.7177 &  25.95 &  0.7670 &  29.02\\ 
    9   & T37 &     0.7912 &  32.39 &  0.8216 &  35.07\\ 
    10   & T27 &     0.8102 &  31.62 &  0.8307 &  32.00\\ 
    11   & T26 &     0.7837 &  26.91 &  0.8362 &  29.61\\ 
    12   & T28 &     0.8498 &  30.94 &  0.8879 &  32.96\\ 
    13   &     &     0.8934 &  35.16 &  0.9265 &  37.10\\ 
    14   & B01 &     0.9062 &  36.33 &  0.9434 &  38.07\\ 
    15   & T03 &     0.9084 &  36.37 &  0.9444 &  38.07\\ 
    16   &     &     0.9193 &  35.78 &  0.9530 &  38.50\\ 
    17   & T09 &     0.9531 &  34.39 &  0.9666 &  34.77\\ 
    18   & B02 &     0.9747 &  39.79 &  0.9724 &  39.54\\ 
    19   & T11 &     0.9836 &  40.91 &  0.9939 &  42.55\\ 
    20   &     &     0.9855 &  40.38 &  0.9945 &  42.98\\ 
    21   & B03 &     0.9827 &  42.16 &  0.9958 &  44.77\\ 
    22   & B04 &     0.9993 &  46.03 &  0.9997 &  48.60\\ 
    23   & T10 &     0.9996 &  45.10 &  1.0000 &  45.50\\ 
    \bottomrule
    \end{tabular}
    \label{tab:PA_results}
\end{table}

\begin{table}[h]
    \centering
    \caption{ASVspoof 2021 progress and evaluation results for DF. Results shown in terms of pooled normalised minimum t-DCF and pooled EER [\%].}
    \begin{tabular}{cccc}
    \toprule
    
           &          &    {Progress set} & {Evaluation set} \\
       \#  & ID       &      EER          &  EER       \\ 
    \midrule
        1   & T23 &     0.24 &  15.64\\ 
        2   & T20 &     1.79 &  16.05\\ 
        3   & T08 &     2.93 &  18.30\\ 
        4   &     &     1.65 &  18.80\\ 
        5   & T06 &     8.29 &  19.01\\ 
        6   & T22 &     2.17 &  19.22\\ 
        7   & T03 &     2.32 &  19.24\\ 
        8   & T01 &     4.72 &  19.70\\ 
        9   & T36 &     0.78 &  20.23\\ 
        10  & T19 &     3.00 &  20.33\\ 
        11  & T31 &     6.88 &  20.33\\ 
        12  &     &     2.25 &  20.63\\ 
        13  &     &     2.00 &  20.82\\ 
        14  & T16 &     1.98 &  20.84\\ 
        15  &     &     4.84 &  21.61\\ 
        16  &     &     5.06 &  21.67\\ 
        17  &     &     2.37 &  22.03\\ 
        18  &     &     6.10 &  22.38\\ 
        19  & B04 &     6.10 &  22.38\\ 
        20  & T25 &     4.18 &  22.62\\ 
        21  & B03 &     11.61 &  23.48\\ 
        22  &     &     0.10 &  23.57\\ 
        23  &     &     5.36 &  23.88\\ 
        24  & T35 &     0.42 &  24.12\\ 
        25  &     &     11.82 &  24.89\\ 
        26  & T30 &     4.66 &  25.21\\ 
        27  & B02 &     21.01 &  25.25\\ 
        28  & T26 &     16.40 &  25.41\\ 
        29  & B01 &     17.63 &  25.56\\ 
        30  &     &     3.40 &  26.67\\ 
        31  & T21 &     17.20 &  28.96\\ 
        32  &     &     9.07 &  29.25\\ 
        33  &     &     29.63 &  29.75\\ 
    \bottomrule
    \end{tabular}
    \label{tab:DF_results}
\end{table}


\end{document}